\documentclass[prd,aps,twocolumn,lengthcheck,showpacs,preprintnumbers,
nofootinbib,letterpaper,amsmath,amssymb,superscriptaddress,nofootinbib]{
revtex4-1}
\pdfoutput=1
\usepackage{color}
\usepackage{graphicx}  
\usepackage{float}
\usepackage{amsmath}
\usepackage{times}
\usepackage{bm}
\usepackage{acronym}
\usepackage{xspace}
\usepackage{subfigure}
\usepackage{enumitem}
\usepackage{array}
\usepackage[hidelinks]{hyperref}

\usepackage{breakurl}

\def\beq{\begin{equation}}
\def\eeq{\end{equation}}
\def\bea{\begin{eqnarray}}
\def\eea{\end{eqnarray}}

\newfont{\cursive}{pzcmi at 9pt}
\def\~t{\tilde{t}}

\begin{document}


\title{Statistical and systematic errors for gravitational-wave inspiral
signals: \\A principal component analysis}
\preprint{AEI-2013-049}
\preprint{LIGO-P1300017}

\newcommand{\AEI}{\affiliation{Max-Planck-Institut f\"ur  
Gravitationsphysik, Callinstrasse 38, D-30167 Hannover, Germany}}

\author{Frank Ohme}
\affiliation{School of Physics and Astronomy, Cardiff University, The Parade,
Cardiff, CF24 3AA, United Kingdom}
\AEI

\author{Alex B. Nielsen}
\AEI

\author{Drew Keppel}
\AEI

\author{Andrew Lundgren}
\AEI

\date{\today}

\begin{abstract} 
Identifying the source parameters from a gravitational-wave measurement alone is
limited by our ability to discriminate signals from different sources and the
accuracy of the waveform family employed in the search. Here we address both
issues in the framework of an adapted coordinate system that allows for linear
Fisher-matrix type calculations of waveform differences that are both accurate
and computationally very efficient. We investigate statistical errors by using
principal component analysis of the post-Newtonian (PN) expansion coefficients,
which is well conditioned despite the Fisher matrix becoming ill conditioned for
larger numbers of parameters.  We identify which combinations of physical
parameters are most effectively measured by gravitational-wave detectors for
systems of neutron stars and black holes with aligned spin. We confirm the
expectation that the dominant parameter of the inspiral waveform is the chirp
mass. The next dominant parameter depends on a combination of the spin and the
symmetric mass ratio.  In addition, we can study the systematic effect of
various spin contributions to the PN phasing within the same parametrization,
showing that the inclusion of spin-orbit corrections up to next-to-leading
order, but not necessarily of spin-spin contributions, is crucial for an
accurate inspiral waveform model. This understanding of the waveform structure
throughout the parameter space is important to set up an efficient search
strategy and correctly interpret future gravitational-wave observations.  %
\end{abstract}

\maketitle

\acrodef{GW}{gravitational-wave}
\acrodef{BH}{black hole}
\acrodef{NS}{neutron star}
\acrodef{PCA}{principal component analysis}
\acrodef{PN}{post-Newtonian}
\acrodef{SNR}{signal-to-noise ratio}

\newcommand{\PN}[0]{\ac{PN}\xspace}
\newcommand{\BH}[0]{\ac{BH}\xspace}
\newcommand{\NS}[0]{\ac{NS}\xspace}
\newcommand{\PCA}[0]{\ac{PCA}\xspace}
\newcommand{\GW}[0]{\ac{GW}\xspace}
\newcommand{\SNR}[0]{\ac{SNR}\xspace}

\section{Introduction}

Ground-based \GW detectors of the Laser Interferometer Gravitational-wave
Observatory (LIGO)~\cite{Abbott:2007kv,Sigg:2008zz,Smith:2009bx,Harry:2010zz}
and Virgo \cite{Acernese:2008zzf,Accadia:2011zz} collaborations are currently
being upgraded to provide sensitivities capable of directly detecting \acp{GW}
from compact binary coalescences of binary \acp{BH} and \NS systems
\cite{Abadie:2010cf}. Such detections would constitute the first direct
detection of \ac{NS}-\BH and binary \BH systems. The gravitational waveforms
from these systems will provide unprecedented information about the physical
nature of these systems, and extracting this information relies on overlapping
the noisy detector data with accurate theoretical signal predictions.

The waveform from a quasicircular inspiraling compact binary system can be
obtained from
knowledge of the energy and energy flux of the system. In general relativity
these can be calculated perturbatively in a $v/c$ expansion (where $v$ is the
relative velocity of the bodies, $c$ is the speed of light), known as a \PN
expansion (see, e.g., \cite{lrr-2006-4} and references therein). These
calculations provide the coefficients in such an expansion in terms of the
fundamental physical parameters. Various different expansion schemes exist that
lead to different \emph{approximants} \cite{Buonanno:2009zt}. For
quasicircular, adiabatic orbits, the tangential velocity $v$ can be related to
the orbital frequency of the compact bodies, which, for the dominant \GW mode,
is equivalent to half the \GW frequency. In this way an expansion in \GW
frequency, $f$, can be obtained, with each successive term corresponding to a
higher order in the \PN expansion.

These waveforms depend on a number of physical parameters such as the masses
and magnitudes and orientations of the objects' spins. An important task will
be extracting as much of this information as possible given the observational
constraints and detector sensitivities.  Although the masses and spins of the
constituent objects are typically the parameters of greatest astrophysical
interest, in practice the detectors are actually sensitive to combinations of
these parameters. This is because rather than variations in the individual
source parameters, only sufficiently strong \emph{waveform
variations} that are louder than the noise background can be
distinguished by the detector. An example of this may be seen already in the
Newtonian regime where the waveform of the binary inspiral depends only on what
is commonly called the chirp mass, ${\cal{M}}$, a combination of the two
individual masses $m_{1}$ and $m_{2}$, given by
${\cal{M}}=(m_{1}m_{2})^{3/5}/(m_{1}+m_{2})^{1/5}$. Systems with the same chirp
mass emit \acp{GW} with the same phase in this simple approximation, and they
cannot be
distinguished. Although this degeneracy is broken by higher order terms, it remains
true that \GW detections can put much stronger constraints on a combination of
the masses characteristic of the binary system, in this case the chirp mass,
than it can on the physical parameters of either individual object.

One major difference between Newtonian dynamics and general relativity is that
in the relativistic domain the spin angular momenta of the inspiraling objects
affects the orbit and thus the gravitational waveform. In general relativity
these spin effects first show up in the \PN expansion as spin-orbit
interactions at 1.5PN order, corresponding to $(v/c)^3$ order corrections. At
higher orders one also encounters spin-spin interactions. With \PN corrections
beyond the Newtonian term and in particular spin effects, it is less obvious
which combinations of the physical parameters are most accurately measured, but
empirical overlap studies \cite{Baird:2012cu,Hannam:2013uu} have recently
pointed out that apart from the chirp mass, there is a close degeneracy between
mass ratio and spins for \acp{BH} with spins aligned with the orbital
angular momentum. (The waveforms of spinning \acp{NS} exhibit additional
degeneracies, e.g., between the \NS spin and the equation-of-state dependent
quadrupole moment, but in this paper we neglect the spin of the \NS as it is
expected to be small in compact binary systems \cite{Bildsten:1992my,
Kochanek:1992wk}.)

Here we formalize the search for well-measurable parameters and degeneracies in
the \PN inspiral waveform. We employ a linearization of waveform differences
equivalent to the Fisher-matrix approximation \cite{Vallisneri:2007ev}, but we
demonstrate that a  convenient higher-dimensional coordinate choice in terms of
the \PN expansion coefficients allows for very accurate, yet computationally
cheap calculations of the waveform (dis)agreement. 

The method we employ is similar to the one used by Tagoshi and Tanaka
\cite{Tanaka:2000xy}, Sathyaprakash and Schutz \cite{Sathyaprakash:2003ua}, Pai
and Arun \cite{Pai:2012mv} and Brown \emph{et al.}\ \cite{Brown:2012qf}. We
write the waveform as a series expansion in frequency space and use the
expansion coefficients as model parameters to construct a Fisher matrix. Using
the eigenvalues and eigenvectors of this Fisher matrix we then determine which
combinations of the expansion coefficients the detector is most sensitive to,
which amounts to finding the principal components of the Fisher matrix. In
contrast to Pai and Arun our focus is determining the best measured
combinations of parameters given aligned spinning general relativity waveforms
and an Advanced LIGO noise curve \cite{advLIGO}. We also discuss implications
of a parameter dependent frequency cutoff.

An example of \PCA of spinning signals for the proposed LISA (Laser
Interferometer Space Antenna) space-based mission was given in
\cite{Sathyaprakash:2003ua}. We, however, consider ground-based detectors,
specifically Advanced LIGO, where the expected \SNR of most detections is rather
low. Then it will be especially important to know which parameters can be
measured since it is unlikely that all physical parameters will be measurable
with reasonable accuracy. The number of principal components with a prescribed
accuracy determined by the \PCA will define an effective dimension of the space
of solutions to be searched \cite{Brown:2012qf}. A solution space with a small
effective dimension will need relatively few templates to be searched, which
speeds up search times \cite{Tanaka:2000xy,Brown:2012qf}. 

Our main aim here is to determine what the principal components represent
physically. This cannot reduce any uncertainty in the measurement of physical
parameters, which is typically large because of the correlation between the
parameters. However, an uncorrelated set of parameters will give more tightly
constrained directions in the likelihood space and also provide a convenient
coordinate system in which to evaluate the overlap of differing waveforms. 

In calculating the principal components we use as much information about the
\PN coefficients of aligned spinning systems as is currently available. The
functional dependence of the \PN coefficients on the physical parameters
dictates how our principal components vary across parameter space. Furthermore,
we investigate what the contribution of various terms in these coefficients are
and how excluding them might affect parameter values through parameter bias.
This helps to show which terms are important in the parameter estimation
problem and gives some indication of how yet-to-be-calculated terms may affect
our results.

This paper is organized as follows. After a general introduction of the
Fisher-matrix approximation and \PCA in Sec.~\ref{sec:PCA_intro}, we specify
the waveform model in Sec.~\ref{sec:wvmodels} and argue by virtue of the
Bauer-Fike theorem that even though the higher-dimensional Fisher matrix may be
ill conditioned, the corresponding principal components with large eigenvalues
can be calculated accurately. In Sec.~\ref{sec:staterr} we demonstrate
explicitly the superiority of our approach over standard Fisher-matrix
estimates in terms of physical parameters, and we extensively analyze the
physical dependence of the first principal components. Section~\ref{Sec:biases}
extends our algorithm to the case of different waveform models, which enables
us to identify crucial contributions to the \PN phasing. We conclude with
Sec.~\ref{sec:conclusion}.

\section{Waveform and methodology} \label{sec:method}

\subsection{Fisher matrix and principal component analysis} \label{sec:PCA_intro}

The fundamental question that is underlying matched-filter searches for
\acp{GW} is \emph{how different is a waveform $h_1$} (the detected signal)
\emph{to another waveform $h_2$} (the template). Assuming a \ac{GW} detector
with noise spectral density $S_n(f)$, the appropriate difference is commonly
defined by
\begin{equation} \label{eq:waveform_diff}
\| h_1 - h_2 \|^2 = \langle h_1 - h_2, h_1 - h_2 \rangle,
\end{equation}
where the inner product reads
\begin{equation} \label{eq:innerprod}
\langle h_1,h_2\rangle = 4\, \mathrm{Re}\int^{f_{\rm max}}_{f_{\rm
min}}\frac{\tilde h_1(f) \, \tilde h_2^\ast (f)}{S_{n}(f)} \, df .
\end{equation}
Here, $\tilde h$ denotes the Fourier transform of $h$ and $^\ast$ is the
complex conjugation. Throughout this paper, we shall assume the noise spectral
density of Advanced LIGO, in the zero detuned high power configuration detailed
in \cite{advLIGO}, with a lower cutoff frequency at $f_{\min} = 15\, \rm{Hz}$.
The upper frequency cutoff, $f_{\max}$, is determined by the signal templates
we assume. Since we are dealing exclusively with inspiral signals and ignoring
the merger and ringdown phase, it is common practice to cut the inspiral
frequencies at the equivalent innermost stable circular orbit of the
Schwarzschild spacetime, at
\begin{equation}
f_{\max}=1/(6^{3/2}\pi M) \label{eq:fISCO} 
\end{equation}
 (where $M$ is the total mass of the system). We could use a more general,
possibly spin-dependent, description of the upper cutoff frequency, but that
is not the focus of this paper, and our algorithm is readily applicable to more
complicated forms of $f_{\max}$.

By evaluating the distance \eqref{eq:waveform_diff} or the overlap
\eqref{eq:innerprod} one can, with some confidence, draw conclusions about the
origin of the detected signal, if the template agrees very well with the data.
Two main issues arise, however. Various templates with different source
parameters may all agree well in terms of the predicted \ac{GW} signals, and
if the remaining small differences are buried below the detector noise level
it becomes impossible to definitively identify the true source
parameters from just the \ac{GW} observation. In addition, the template family
may not be an exact representation of the real waveform, which again limits our
ability to unambiguously identify the source of the detected signal. 

We shall analyze both effects here, \emph{statistical} uncertainties due to
similar waveforms for different parameters and \emph{systematic} biases due to
inaccuracies in the waveform model. As such analyses are crucial for the
correct interpretation of \ac{GW} observations, there are already a number of
publications addressing these issues under various assumptions. In particular,
\PN inspiral waveforms have been analyzed by several authors, who calculated
the measurement accuracy of various source parameters assuming a similar
frequency-domain model to the one we employ here, but to lower \PN expansion
order \cite{Finn:1992xs, Jaranowski:1994xd, Cutler:1994ys, Poisson:1995ef,
Nielsen:2012sb} or only for nonspinning systems \cite{Arun:2004hn}. Systematic
errors between different \PN approximants describing nonspinning systems have
been studied with extensive overlap calculations in \cite{Buonanno:2009zt}.

We elaborate on the existing insights here by improving the linearization of
\eqref{eq:waveform_diff} through a suitable coordinate choice in combination
with a \ac{PCA}, which allows us to understand in a systematic way which
combinations of physical parameters are best constrained and which analytical
contributions to the inspiral waveform are crucial to correctly recover the
source parameters. 

The details of our strategy are as follows. We assume that $h_1$ and $h_2$ can
be parametrized by a common waveform manifold $h$ such that $h_1 = h(\theta)$
and $h_2 = h(\hat \theta)$, where $\theta$ is the vector of waveform parameters
with components $\theta_i$ (we shall put these in concrete terms in
Sec.~\ref{sec:wvmodels}). With a minimization of the distance
\eqref{eq:waveform_diff} in mind, we next apply the well-known linearization 
\begin{equation} \label{eq:waveform_diff_lin}
\| h(\theta) - h(\hat \theta) \|^2 \approx \sum_{i,j} \Gamma_{ij} \, \Delta
\theta_i \, \Delta \theta_j ,
\end{equation}
where $\Delta \theta = \theta -\hat \theta$ and $\Gamma_{ij}$ is the Fisher
information matrix,
\begin{equation} \label{eq:Fisher} 
\Gamma_{ij} = \left\langle\frac{\partial h}{\partial \theta_i} , \frac{\partial
h}{\partial \theta_j} \right\rangle .
\end{equation}
For more details about this approach and the validity of the Fisher-matrix
approximation, see for instance \cite{Vallisneri:2007ev}. We simply use it as a
convenient linearization here.

The inverse of the Fisher matrix is the covariance matrix of the waveform
parameters, $\mathcal C = \Gamma^{-1}$, and instead of quoting the variances of
the used parameters (as
done, e.g., in \cite{Finn:1992xs, Jaranowski:1994xd, Cutler:1994ys,
Poisson:1995ef, Arun:2004hn, Nielsen:2012sb}), we proceed by diagonalizing the
Fisher matrix. The result can be written as
\begin{equation}
\Gamma_{ij} = \sum_{k,l} \Lambda^T_{ik} \; \Sigma_{kl} \; \Lambda_{lj} ,
\end{equation}
where $\Lambda_{ij}$ denotes the $j$th component of the $i$th eigenvector of
the Fisher matrix. $\Sigma_{ij}$ is a diagonal matrix with positive eigenvalues
on the diagonal, i.e., $\Sigma_{ij} = \lambda_i \, \delta_{ij}$. Since the
eigenvectors are also eigenvectors of the covariance matrix, $\mathcal C$, we
have thus
performed a \ac{PCA}, and the vector
\begin{equation} \label{eq:mu_definition}
\mu_i =  \sum_j \Lambda_{ij} \, \theta_j
\end{equation}
describes the \emph{principal components} of the system.

Working with these coordinates rather than the original parametrization has the
great advantage that the waveform difference \eqref{eq:waveform_diff_lin}
becomes simply
\begin{equation} \label{eq:wfdiff_eigenvec}
\| h(\theta) - h(\hat \theta) \|^2 = \sum_i \lambda_i \, \Delta \mu_i^2 .
\end{equation}
We can now easily conclude from the size of the eigenvalues which principal
components (or which principal directions in parameter space) affect the
waveform strongly. This will be important to understand how well constrained
and therefore measurable certain parameter combinations are, given that
waveforms that differ below the noise floor cannot be distinguished from each
other.

Typically, the smallest difference that is measurable is quoted to be 
\begin{equation}
\| h(\theta) - h(\hat \theta) \|^2 = 1, 
\end{equation}
see \cite{Lindblom:2008cm} and further discussions in \cite{Lindblom:2009ux,
Damour:2010zb}. Here, however, we follow the recent discussion by Baird
\emph{et al.}\ \cite{Baird:2012cu}, who detail that the 90\% confidence
interval in the posterior probability distribution is given to linear order by
\begin{equation} \label{eq:confidenceInterval}
\| h(\theta) - h(\hat \theta) \|^2 < \chi_k^2,
\end{equation}
where $\chi_k^2$ is the $\chi^2$ value for which the probability of obtaining
that value or less in a $\chi^2$ distribution with $k$ degrees of freedom is
90\%. We shall later restrict ourselves to three physical parameters (the two
masses of the compact objects and one spin magnitude) where $\chi_3^2 \approx
6.25$ and waveforms with distance
\begin{equation} \label{eq:90percInterval}
\| h(\theta) - h(\hat \theta) \|^2 < 6.25
\end{equation}
cannot be distinguished at 90\% confidence. Note that for \SNR~10, this is
approximately equivalent to the region of waveforms with overlap greater than
97\%~\cite{Baird:2012cu}.

\subsection{Waveform model} \label{sec:wvmodels}

Let us specify in this section what waveform manifold $h(\theta)$ we are using,
which set of parameters $\theta_i$ we are considering, and how we can take
advantage of the methods presented in the previous section.

In \ac{GW} data analysis, the inspiral waveform of a coalescing compact binary
is most conveniently expressed in a closed form in the Fourier domain, which
allows for a very fast evaluation of thousands of templates, if needed. We
shall use the same signal description here, which is commonly referred to as
the ``TaylorF2'' approximant. Derived from the stationary-phase approximation
of the \ac{PN} energy balance law~\cite{Damour:2000zb, Damour:2002kr}, it reads
\begin{equation} \label{eq:h_definition}
\tilde h(f) = \mathcal A \left( \frac{f}{f_0} \right)^{\! -7/6} e^{i\psi(f)},
\end{equation}
where $\mathcal A$ is an amplitude term which we set by requiring a particular
\ac{SNR}, $\langle h, h \rangle = \rho^2$, and $f_0$ is an arbitrary reference
frequency as detailed below. We do not consider contributions
from higher harmonics, which can be found in~\cite{Blanchet:2008je,
Arun:2008kb}; hence $\mathcal A$ is simply a constant determined by the
binary's total mass, distance, orientation and sky location. With
this assumption
\eqref{eq:h_definition} is often called the \emph{restricted} \ac{PN}
waveform~\cite{Buonanno:2009zt}. 

The phase, $\psi(f)$, is given as a series in the gravitational wave frequency
$f$, 
\begin{equation} \label {eq:PNexpancoeffs}
\psi(f) = \sum_{k=0}^N \left( \frac{f}{f_0} \right)^{(k-5)/3}  \left[ \psi_k +
\psi^{\log}_k \, \log(f/f_0) \right],
\end{equation}
where we introduce $f_0$ to
make all
coefficients dimensionless.  The expansion coefficients $\psi_k$ and
$\psi_k^{\log}$ have been determined within the \ac{PN} formalism in various
publications (see~\cite{lrr-2006-4} and references therein).  Currently, the
highest \PN order to which they are known is 3.5 ($k=7$) for nonspinning
contributions. Spin-dependent contributions enter as leading-order and
next-to-leading order spin-orbit effects at 1.5 and 2.5\PN order ($k=3,
5$)~\cite{Kidder:1995zr, Apostolatos:1994mx, Blanchet:2006gy}. We also include
the tail-induced spin-orbit contribution to the flux at 3\PN order
($k=6$)~\cite{Blanchet:2011zv}. In addition, spin-spin effects are included at
relative 2\PN order~\cite{Kidder:1995zr, Damour:2001tu, Poisson:1997ha,
Mikoczi:2005dn} and a 2.5\PN contribution to the flux is associated with energy
flux into the \ac{BH} \cite{Alvi:2001mx}. The explicit set of coefficients we
employ can be read off the phase expansion in Eq.~(2.91) of~\cite{FrankThesis},
or equivalently Eq.~(A21) of~\cite{Brown:2007jx}, with the exception of the 3PN
term, where we also add the recently calculated tail-induced spin-orbit
term (see Eq.~(6.6) of~\cite{Blanchet:2011zv}). We note that in the course of
finalizing this paper, \PN spin contributions at even higher expansion order
have been computed in \cite{Bohe:2013cla} which are not included in this study.

Three \emph{physical} source parameters define the binaries we consider: one
object is a \ac{BH} of mass $m_1$, the other is a \ac{NS} of mass $m_2$. In
addition, the \ac{BH} is allowed to have a spin $\bm S_{1}$ aligned with the
direction of the orbital angular momentum $\hat {\bm L}$ of the binary, which
we parametrize through the dimensionless quantity
\begin{equation}
\chi_1 = \frac{\hat{\bm  L} \cdot \bm S_{1}}{m_{1}^2} ~.
\end{equation}
The \ac{NS} spin is assumed to be negligible and set to zero. We could also
include the spin of the second compact object without any modification to our
algorithm, as long as the spins are aligned with the orbital angular momentum.
However, astrophysical expectations are that \acp{NS} in compact binaries do
not spin rapidly~\cite{Bildsten:1992my, Kochanek:1992wk}, which is confirmed by
the fact that the highest \NS spin parameter in a binary observed to
date~\cite{Burgay:2003jj} has a value of $\chi \sim 0.05$.  In addition, it was
argued from a purely \GW data analysis point of view that two spin parameters
can efficiently be mapped onto one \emph{effective spin} parameter, essentially
without changing the waveform manifold~\cite{Ajith:2009bn, Santamaria:2010yb,
Ajith:2011ec}. In that sense, one spinning \BH can simply be seen as a
representative of the class of aligned-spin systems with the same effective
spin.

We thus consider the PN coefficients in \eqref{eq:PNexpancoeffs} generally as
functions of $(m_1, m_2, \chi_1)$. Note, however, that we work under the
assumption of general relativity, in which $\psi_1$ and many $\log$ terms are
exactly zero.%
\footnote{In non-Einstein theories some of these terms may not be zero, for
example in Brans-Dicke gravity a term arises already at a value of $k=-2$
\cite{Yunes:2009ke}. In order to capture waveforms of any gravitational theory,
as performed in for example \cite{Li:2011cg}, who include a possible nonzero
$k=1$ term, these terms would need to be incorporated. In this work we focus on
the general relativity results and set these terms explicitly to zero.} 

Two coefficients deserve further attention. $\psi_5$ is a constant phase that
comes with no frequency-dependent factor, and it represents what is usually
referred to as an additional parameter: the ``phase at coalescence'' or the
initial phase. The corresponding initial time or ``time at coalescence'' is
included in $\psi_8$, which is a linear contribution to the phase. When we
estimate waveform differences later, we are not interested in any discrepancy
caused by time or phase shifts, so we shall project these parameters out of the
Fisher matrix.

Now that we have defined our parameters and waveform model, we could proceed by
calculating the Fisher matrix \eqref{eq:Fisher} for the parameters $(m_{1},
m_{2}, \chi_1, \phi_0, t_0)$, where $\phi_0$ and $t_0$ represent the free time
and phase shift, or equivalently for any other five-dimensional parametrization
that can uniquely be mapped to the above parameters. It was found, however, in
various publications that assuming a moderate \ac{SNR} (between 10 and 20) and
the noise spectral density of the (initial or advanced) LIGO detector leads to
rather large statistical parameter biases in some parameters of interest. The
linearization of the waveform difference implies that waveforms can
definitively be told apart only if parameters like the individual masses or the
mass ratio are considerably different \cite{Cutler:1994ys, Poisson:1995ef,
Nielsen:2012sb} (to the order of tens of percents and more). The validity of
the linearization is rightly questioned in these cases, and  the true
confidence interval \eqref{eq:confidenceInterval} likely has more structure
than the ellipsoid predicted in the Fisher-matrix approach.

We partly circumvent these problems here by using the \PN coefficients
themselves as free parameters rather than the original physical parameters,
i.e.,
\begin{equation} \label{eq:PNcoeffpar}
\theta_i = \Big \{ \psi_k \Big \} \cup \Big \{ \psi_k^{\log} \Big \}.
\end{equation}
All nonzero \PN coefficients up to $N= 8$ are now considered to be free
parameters, which yields a ten-dimensional parameter space that includes eight
$\psi_k$ ($k=0, 2, 3, \ldots, 8$) and two $\psi_k^{\log}$ ($k = 5, 6$).  As
noted above, the same idea, combined with an eigenvector analysis, has already
been presented by Tagoshi and Tanaka~\cite{Tanaka:2000xy} and Brown \emph{et
al.}~\cite{Brown:2012qf} in the context of template placing, and by Pai and
Arun~\cite{Pai:2012mv} to probe \PN theory with \ac{GW} observations. We shall
show below how this trick is also useful to understand statistical and
systematic errors of inspiral waveforms. One important feature we will exploit
is that the Fisher matrix \eqref{eq:Fisher} becomes almost
parameter independent for the choice \eqref{eq:PNcoeffpar}, which makes
extrapolating the waveform difference \eqref{eq:waveform_diff_lin} to large
parameter variations much more accurate~\cite{Tanaka:2000xy, Babak:2006ty}.
Specifically, the only parameter dependence in the resulting matrix
\begin{equation} \label{eq:Fisher_parindependent}
\Gamma_{ij}  =  \left \vert \mathcal A \right \vert^2
\int_{f_{\min}}^{f_{\max}} \left( \frac{f}{f_0} \right)^{\!\kappa}
\frac{\log^\delta (f/f_0)}{S_n(f)} \, df
\end{equation}
is inherited from the upper cutoff frequency \eqref{eq:fISCO}. The exponents
$\kappa$ and $\delta$ are solely functions of $i$ and $j$. ($\delta$ is simply 0, 1 or 2, depending on the number of logarithmic
coefficients in $\{ \theta_i$, $\theta_j\}$; $\kappa$ depends on the \PN
order that each $\bm \theta$ component corresponds to. Given the mapping
$i \mapsto k(i)$, we can express $\kappa = [k(i) + k(j) -17]/3$.)

\begin{figure}
\centering
 \includegraphics[width=0.4\textwidth]{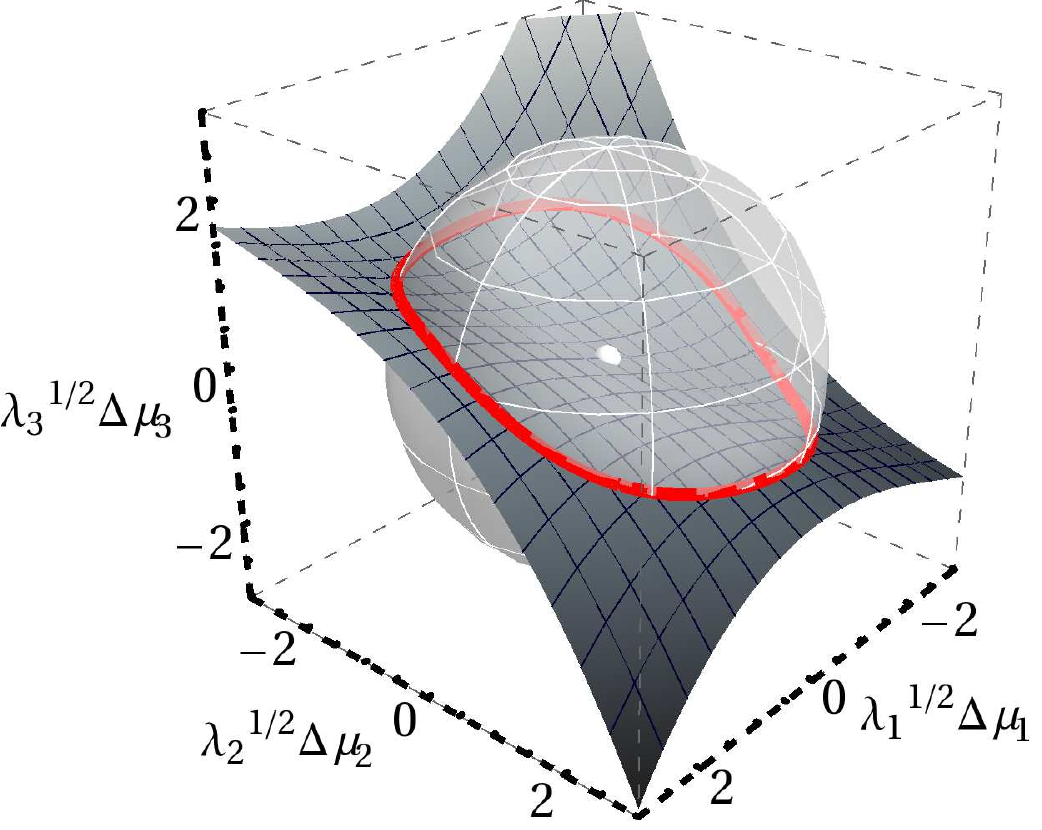}
 \caption{Geometric interpretation of calculating waveform differences in a
higher-dimensional flat space \eqref{eq:wfdiff_eigenvec}. The actual manifold
of inspiral waveforms is illustrated as a two-dimensional curved surface that
can be embedded in a higher-dimensional flat space [see coordinate choice
\eqref{eq:PNcoeffpar}], here depicted in three dimensions. The confidence
interval has a trivial geometry in the higher-dimensional space (here
illustrated as a ball around the target waveform, which is shown as a white
dot), which has to be projected back onto the physical waveform manifold.}
 \label{fig:geometry}
\end{figure}

We shall later show how we can exploit the fact that in this convenient
parametrization, finding the confidence interval around a given target signal
transforms to a simple geometric exercise in the flat space, assuming we can fix
the cutoff frequency parameter independently (which we shall do in
Secs.~\ref{sec:PCA_interpretation} and \ref{Sec:biases}). An illustration of
advantages and caveats of this geometric interpretation is provided by
Fig.~\ref{fig:geometry}. In addition, we shall show in
Sec.~\ref{sec:diffparametrizations} how a weak nonflatness, inherited by a
parameter dependent $f_{\max}$, can be incorporated properly in accurate overlap
calculations.

As mentioned earlier, the waveform variations we are interested in are obtained
by projecting the time and phase shifts out of the Fisher matrix, and we can do
so following the simple procedure
\cite{Owen:1995tm}
\begin{equation} \label{eq:Fisher_projected}
\tilde \Gamma_{ij} = \Gamma_{ij} - \sum_{a,b} \Gamma_{ia} \gamma^{ab}
\Gamma_{bj}.
\end{equation}
Here, $\gamma^{ab}$ denotes the inverse of the two-dimensional submatrix of
$\Gamma$ corresponding to the parameters $\psi_5$ (phase shift) and $\psi_8$
(time shift). The projected matrix $\tilde \Gamma_{ij}$ is effectively
eight dimensional, thus we shall find eight eigenvectors and eigenvalues that
govern waveform changes according to \eqref{eq:wfdiff_eigenvec}.

Let us mention two caveats before we continue with presenting our results.
First, in calculating the Fisher matrix with more than just our five physical
parameters, we implicitly treat all \PN coefficients as independent, which is
clearly not true for the underlying lower-dimensional waveform manifold. So we
have to take care to  ``project'' our results back onto physically meaningful
quantities. We shall do so by reexpressing the principal components not just
as functions of \PN coefficients, but eventually as functions of the physical
parameters.

Doing so will render all the principal components as functions of the physical
parameters and in the underlying lower-dimensional waveform manifold they will
not all be independent. When the number of \PN coefficients is larger than the
number of physical parameters (which it typically will be in our applications)
the excess principal components will not give additional information about the
physical parameters. In practice we will find that the errors on these excess
principal components will be large anyway and they can be ignored. Note that a
different strategy was followed in \cite{Brown:2012qf} where the authors perform
an additional \PCA to rotate the original principal directions onto the physical
manifold. This additional step, however, depends on a freely chosen sample of
source parameters, and we instead focus on the identification of, in absolute
terms, best measurable combinations of \PN coefficients, independent of how
the particular model is oriented in that space.

\subsection{Numerical stability}

The second issue to be aware of is that the condition number of the Fisher
matrix, which we can view
as the ratio of greatest and smallest eigenvalue of $\tilde \Gamma_{ij}$, can
become quite large for the matrices we are considering. This means that
inverting the matrix can become susceptible to numerical errors. Calculating the
eigenvalues for these matrices, however, is still a well-conditioned problem by
virtue of the Bauer-Fike theorem which states \cite{datta2010numerical}
\begin{equation} \label{eq:eigenval_variation}
|\lambda_{\Gamma} - \lambda_{\Gamma+\delta\Gamma}| \leq \kappa_{2}(\Lambda)
\left \| \delta\Gamma \right \|_{2} 
\end{equation}
for a diagonalizable matrix $\Gamma$ with eigenvalue $\lambda_{\Gamma}$ and a
perturbed matrix $\Gamma+\delta\Gamma$ with eigenvalue
$\lambda_{\Gamma+\delta\Gamma}$. The factor $\kappa_{2}$ is the condition number
of the diagonalizing matrix $\Lambda$ and $\| \delta\Gamma \|_2$ denotes the
matrix 2-norm of the perturbation matrix $\delta\Gamma$.  Since in the case of a
real symmetric matrix $\Gamma$, the diagonalizing matrix $\Lambda$ is orthogonal
with condition number $\kappa_{2}(\Lambda)=1$, we are guaranteed that the
eigenvalue problem is well conditioned. A small change in $\Gamma$ caused by
numerical truncation error will only cause a small change in the eigenvalues.
This well-conditioned property is in fact true for any normal matrix $A$
satisfying $A^{\dagger}A=AA^{\dagger}$ with well-separated eigenvalues. This
means that the large eigenvalues (and this is what we are interested in) of the
Fisher matrix can be reliably computed even when the original matrix contains
many PN coefficients and is itself ill conditioned.

There are, of course, more sources of error in addition to the numerical
inversion of a badly conditioned matrix. In our case of the eight-dimensional
matrix
$\tilde \Gamma$ (as introduced in Sec.~\ref{sec:wvmodels}) we find the largest
source of error to be the numerical integration of
(\ref{eq:Fisher_parindependent}). We compared various methods -- simple uniform
discretization with subsequent extrapolation to infinite resolution and local
adaptive methods -- and the difference we find is of the order $\| \delta \tilde
\Gamma \|_2 \sim 10^{-2}$ for the target signal we shall analyze in the next
section (see Fig.~\ref{fig:compFisherMethods} and
Table~\ref{tab:PNcoefferrors}). We have to compare this number to the size of
the eigenvalues that we try to calculate, and by doing so we shall find in
Table~\ref{tab:PNcoefferrors} that the first (i.e., greatest) three eigenvalues
are guaranteed to be well determined, subsequent eigenvalues are of the same
order or smaller than the error bound (\ref{eq:eigenval_variation}). Note,
however, that by comparing the eigenvalues and eigenvectors obtained from
different integration techniques directly, we find that the uncertainty in the
fourth eigenvector is still negligible. The numerical values of higher
eigenvalues cannot be trusted, but we shall show that this does not harm our
calculations.  

There is a similar estimate for the eigenvectors $\bm x_{i}$ that are perturbed
by $\delta \bm x_{i}$ under a perturbation of the Fisher matrix by
$\delta\Gamma$. In this case we have \cite{datta2010numerical}
\begin{equation} \label{eq:eigenvec_variation}
\delta \bm x_{i} = \sum_{j \neq i} \frac{\bm x_{j}^{T} \, \delta \Gamma \, \bm
x_{i}}{(\lambda_{i}-\lambda_{j})} \bm x_{j} + {\mathcal{O}}(\| \delta\Gamma
\|^{2}) .
\end{equation}
For the eigenvectors with large eigenvalues $\lambda_{i}$, this ensures that
their components are also well conditioned despite the matrix $\Gamma$ being
ill conditioned. This is sufficient for our purposes since we focus on waveform
differences calculated via \eqref{eq:wfdiff_eigenvec}, which are dominated by
the eigenvectors with large eigenvalues. In numerical testing it was observed
that the large eigenvalues are always well separated and that the eigenvalues
and eigenvector components were stable for small changes of the Fisher matrix
$\Gamma$. This stability of the diagonalization process applies to both the
numerical errors and also to small deviations in the noise spectral density,
which will also give rise to small changes in the Fisher matrix.

\section{Measurable parameter combinations -- Statistical error}
\label{sec:staterr}

Let us now present the results we obtain with the method and waveform model
introduced in Sec.~\ref{sec:method}. We first estimate statistical errors by
asking the question: \emph{Which waveforms in a neighborhood of a given
signal are similar to the latter to an extent that they cannot be distinguished
at a 90\% confidence level?} We shall show that
\begin{enumerate}[label=(\arabic*)]
 \item our method of estimating waveform differences is superior to standard
Fisher-matrix estimates (that are carried out in terms of the physical
parameters), and we find no considerable differences between our
computationally cheap method and full overlap calculations;
 \item because of an approximate degeneracy between mass ratio and spin, the
individual masses cannot be determined accurately in \ac{GW} observations
alone, but
 \item through a \ac{PCA} we are able to identify the parameter combinations
that are accurately measurable.
\end{enumerate}
Our results complement previous publications that estimated the measurability
of source parameters of spinning systems either by Fisher-matrix
calculations~\cite{Cutler:1994ys, Poisson:1995ef, Nielsen:2012sb} or recently
by direct overlap calculations~\cite{Baird:2012cu, Hannam:2013uu}.

\subsection{Advantage of different parametrizations}
\label{sec:diffparametrizations}

The target system we consider for illustration is a binary containing a
nonspinning \ac{NS} of mass $1.35M_\odot$ and a \ac{BH} with $5M_\odot$ and a
spin of $\chi_1 = 0.3$. We further assume a moderately high \ac{SNR} of 20.
We can now easily demonstrate the efficacy of our approach by estimating the
90\% confidence interval [defined by \eqref{eq:confidenceInterval}] around the
target signal with various strategies and compare the results with proper
overlap calculations. 

The standard Fisher-matrix estimate in terms of $\theta_i
= \{\log M, \log \eta, \chi_1, t_0, \phi_0 \}$ is the simplest way of obtaining
a
multidimensional ellipse around the target parameters. Here we adopt the
commonly used parametrization in terms of the symmetric mass ratio $\eta$ and
the total mass $M$, 
\begin{equation} \label{eq:symm_massratio_M}
\eta = \frac{m_1 \, m_2}{M^2}, \qquad M = m_1 + m_2~.
\end{equation}
Using the logarithms of the mass-dependent parameters improves the condition
number of the Fisher matrix, and the square root of the diagonal elements of
the inverse matrix (properly scaled) immediately yield the dimensions of the
confidence interval, 
\begin{align} \label{eq:standardFisher_errors}
\frac{\Delta M}M &\approx 60\% , & \frac{\Delta \eta}\eta &\approx 100 \%, &
\Delta \chi_1 &\approx  0.4 \nonumber \\ 
\Delta t_0 &\approx 10\, {\rm ms}, & \Delta \phi_0 &\approx 52\, {\rm rad}.
\end{align}
Evidently, these ranges are extremely large, and we would have to incorporate
prior restrictions of the parameters to obtain a slightly more realistic
estimate of the parameter uncertainties \cite{Cutler:1994ys, Poisson:1995ef,
Vallisneri:2007ev, Nielsen:2012sb}.
However, we merely use it as an illustration here. 

Of course, it is well known that particular parameter combinations can
potentially be measured much more accurately. The canonical example is the
chirp mass
\begin{equation} \label{eq:Mchirp}
\mathcal M = M \eta^{3/5}
\end{equation}
which governs the Newtonian-order \PN phase coefficient. In the above example,
Fisher-matrix calculations in terms of $\mathcal M$ instead of $M$ (the other
parameters remain unchanged) reveal
\begin{equation} \label{eq:MchirpError}
\frac{\Delta \mathcal M}{\mathcal M} \approx 0.32\%,
\end{equation}
and we shall later formalize the search for such well-determined parameters.

\begin{figure}
\centering
 \includegraphics[width=0.42\textwidth]{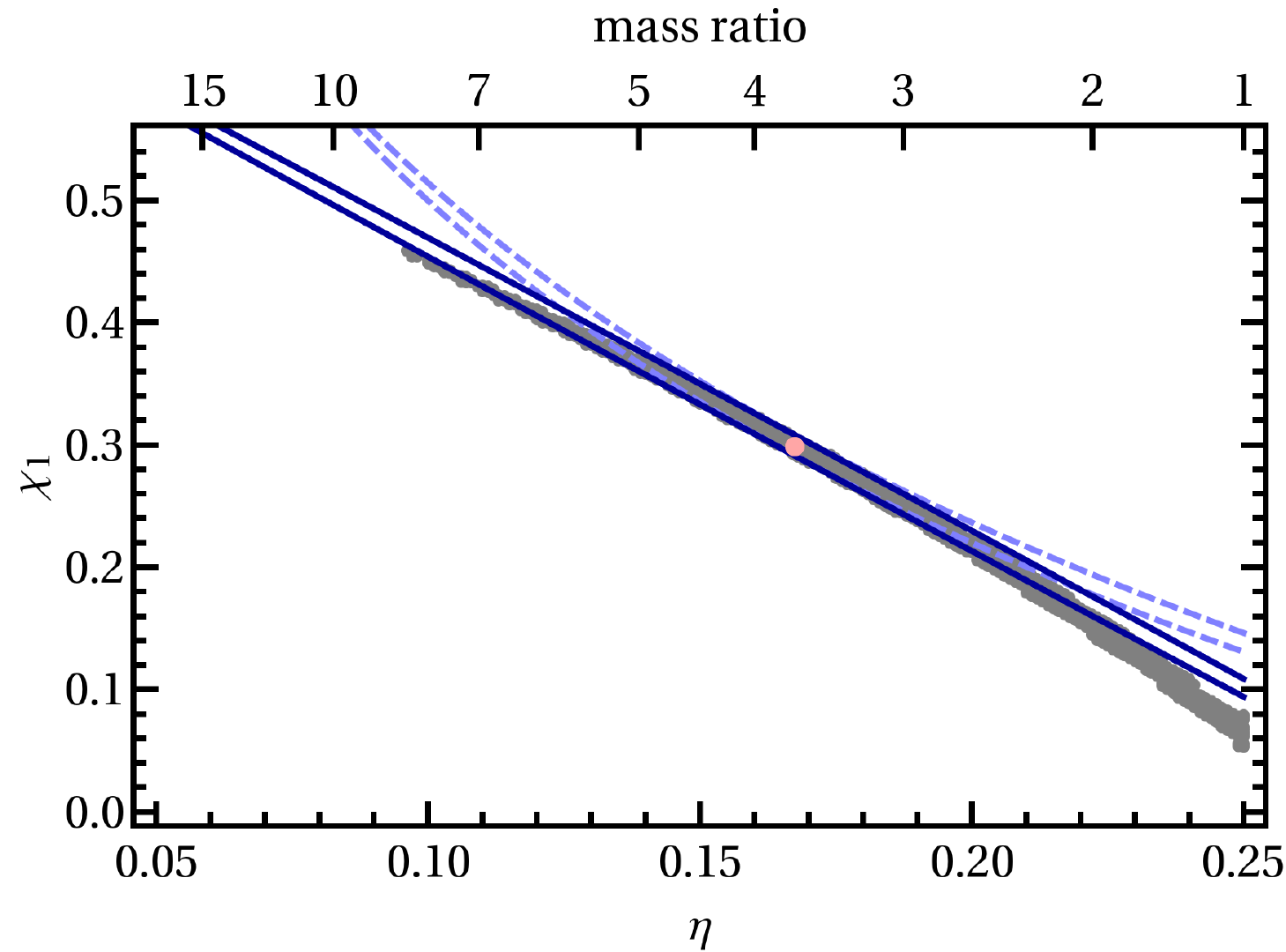} 
 \caption{The estimated 90\% confidence interval around a \ac{NS}-\ac{BH}
signal with component masses $1.35M_\odot$ and $5M_\odot$, SNR 20 and a \BH spin
of $\chi_1 = 0.3$, as indicated by the (red online) dot.
Dashed and solid lines are Fisher-matrix estimates in terms of the physical
parameters (with parameter differences in terms of the logarithm of parameters,
e.g. $\Delta \log \eta$ for the dashed lines or with parameters
linearized as $\Delta \eta/\eta$ for the solid lines). The gray region is
obtained by
actual waveform overlaps, or equivalently by the \ac{PCA} introduced in
Sec.~\ref{sec:method}. Both methods yield indistinguishable results.}
 \label{fig:compFisherMethods}
\end{figure}

For now, we focus on the other physical parameters and show in
Fig.~\ref{fig:compFisherMethods} the boundaries of the confidence interval,
projected onto the $\eta$-$\chi_1$ plane according to
\eqref{eq:Fisher_projected}. As we use $\log \eta$ as one parameter, the
corresponding variation reads $\Delta \log \eta = \log \eta - \log \hat \eta$,
which gives rise to the dashed curve in Fig.~\ref{fig:compFisherMethods}. Since
we work only to linear order in the parameters, we may as well express $\Delta
\log \eta \approx \Delta \eta/\eta = (\eta - \hat \eta)/\eta$, which in turn
leads to the solid ellipse in Fig.~\ref{fig:compFisherMethods}. Already these
two linear approximations differ considerably in the range where we apply them,
indicating that we should not trust the linearization for such high parameter
variations.

Instead, the gray shaded region in Fig.~\ref{fig:compFisherMethods} can be
obtained by actual overlap calculations according to \eqref{eq:waveform_diff}
and \eqref{eq:innerprod}, and then simply recording the area where the distance
between target and model waveform satisfies \eqref{eq:90percInterval}. Note
that in order to compare with the projected Fisher matrix, we optimized the
inner product over all parameters not shown in Fig.~\ref{fig:compFisherMethods}
(these are $t_0$, $\phi_0$ and $M$). This approach, however, entails nontrivial
computational efforts, and a much more efficient method is to use the
parametrization \eqref{eq:PNcoeffpar} discussed in Sec.~\ref{sec:wvmodels}
that embeds the waveform manifold in flat space where the Fisher matrix is
parameter
independent. In that case, calculating the waveform difference
\eqref{eq:waveform_diff_lin} becomes a simple, yet very accurate matrix
multiplication. Indeed, when we calculate the data for
Fig.~\ref{fig:compFisherMethods}, there is no distinguishable difference
between actual overlap results and Fisher-matrix estimates in terms of \PN
coefficients \eqref{eq:PNcoeffpar}, but the computation times differ by several
orders of magnitude, the latter calculation being much faster.\footnote{The
actual improvement in computational cost depends on the details of the
implementation. We tested our speed-optimized single-core implementations of
both Fourier-transform and Fisher-matrix based algorithms, and the latter
performed between 140 and several hundred times faster than the standard
fast-Fourier-transform method, depending on the accuracy of the time
optimization.} 

The details of this fast algorithm are as follows. We calculate and diagonalize
$\Gamma_{ij}$ in terms of the \PN coefficients \eqref{eq:PNcoeffpar}. Inverting
this matrix to estimate the accuracy of these parameters is likely to introduce
large numerical errors, because as noted in both \cite{Pai:2012mv} and
\cite{Berti:2004bd}, large Fisher matrices of this form are ill conditioned and
numerical inversion routines cannot be trusted. However, diagonalizing the
Fisher matrix is numerically more stable due to \eqref{eq:eigenval_variation}
and \eqref{eq:eigenvec_variation}, and we only need to accurately calculate the
eigenvectors with large eigenvalues. 

\begin{table}
\begin{ruledtabular}
 \begin{tabular}{>{$}r<{$} >{$}r<{$} >{$}r<{$} >{$}r<{$} >{$}r<{$}}
  i &  \lambda_i & \Delta \mu_i ~{\rm  (Fisher)} & \Delta \mu_i ~{\rm  (actual)}
& \lambda_i (\Delta \mu_i)^2 ~{\rm  (actual)}\\ \hline 
   1 & 45300 & 0.012 & 0.011 & 5.56 \\
   2 & 80 & 0.28 & 0.27 & 5.73 \\
    3 & 0.84 &2.7 & 1.2 & 1.17 \\
    4 & 0.008 & 28 & 16 & 2.08 \\
    5 & 4 \times 10^{-5}  & 390 & 1.6 &  \sim 10^{- 4} \\
    6 & 4 \times 10^{-8} &  1 \times 10^4 & 46 & \sim 10^{- 4} \\
    7 & 2 \times 10^{-10} &  2 \times 10^5 & 4.9 & \sim 10^{-9} \\
    8 & 1 \times 10^{-13}  & 7 \times 10^6 & 41 & \sim 10^{-10}
 \end{tabular}
\end{ruledtabular}
\caption{The principal components of the Fisher matrix that treats the \PN
phase coefficients as free parameters (see Sec.~\ref{sec:wvmodels}). The target
signal is the same as in Fig.~\ref{fig:compFisherMethods}. We report the
eigenvalues $\lambda_i$ and the theoretical spread $\Delta \mu_i$ (Fisher) in
the 90\% confidence interval, assuming all \PN coefficients to be free and
independent parameters. The latter should be contrasted with the actual
variation $\Delta \mu_i$ (actual) on the lower dimensional waveform manifold,
which in turn affects the waveform difference \eqref{eq:wfdiff_eigenvec} by the
amount stated in the last column.  Note that the numbers shown here depend
on the reference frequency $f_0$ in \eqref{eq:PNexpancoeffs}, and we employed
$f_0 = 200\,{\rm Hz}$.}
\label{tab:PNcoefferrors}
\end{table}

Table~\ref{tab:PNcoefferrors} reports these eigenvalues $\lambda_i$ that enter
the waveform difference through \eqref{eq:wfdiff_eigenvec}. Assuming a
maximally allowed  square difference of $6.25$, Eq.~\eqref{eq:90percInterval},
we can then simply scale the inverse square root of $\lambda_i$ to obtain the
theoretical range of principal components in the confidence interval [denoted
by $\Delta \mu_i$ (Fisher) in Table~\ref{tab:PNcoefferrors}]. However, this
will be of little value, since we cannot easily transform this range back to
physical parameters, and the actual waveform manifold is only a subset of the
eight-dimensional ellipse we have just calculated, see Fig.~\ref{fig:geometry}.

Instead, we  densely populate the (physical) parameter space around the target
signal, transform these coordinates to the $\mu_i$ space (by a matrix
multiplication) and select all points that fulfill \eqref{eq:90percInterval}.
This is a computationally very cheap procedure which allows us to find the
actual spread in both physical parameters and principal components. Of course,
$\Delta \mu_i$, restricted to the physical waveform manifold (labeled by the
word ``actual'' in Table~\ref{tab:PNcoefferrors}), has to be less than or equal
to
the theoretical prediction that assumes all \PN coefficients to be independent,
and indeed, this is what we find in Table~\ref{tab:PNcoefferrors}. Moreover, we
conclude from these numbers that only the first four principal components
contribute significantly to the waveform difference, and we can neglect the
others for all practical purposes, as their actual variation is diminished by
the small associated eigenvalue.  

The superiority of our \PCA over standard Fisher-matrix estimates is based on
two key modifications. First, we increase the dimensionality of our coordinate
space such that the physically allowed waveform manifold is embedded into a
space with only weakly varying matrix coefficients. As stated before, this makes
the extrapolation to large parameter deviations much more accurate. On the other
hand, one might think about using the first principal components as new
coordinates (instead of the physical coordinates $M$, $\eta$ and $\chi_1$)
without increasing the dimensionality, and although locally this is just a
linear transformation, the predicted confidence interval would still be more
accurate. The reason for this is that different coordinate choices yield the
same result locally (i.e., to linear order, as one can also see in
Fig.~\ref{fig:compFisherMethods}), but for larger distances, it becomes
increasingly important which specific set of parameters is bounded by the
Fisher-matrix estimate. Thus, inverting the $\Delta \mu_i$ whose functional form
is adapted to the \PN waveform structure is more accurate than the simple linear
approximation in physical parameters.

Let us make a final remark on the power of our approximation. Previous uses of
\PN coefficients as free parameters in the Fisher matrix have largely neglected
a parameter-dependent cutoff frequency in the inner product
\eqref{eq:innerprod}, mainly because the considered systems had low enough
total mass to place $f_{\max}$ out of the detector's sensitivity band. For
mixed \ac{NS}-\ac{BH} binaries we still may want to neglect the merger and
ringdown of the signal, but the waveform then has to be terminated in the
detector band with a consistent cutoff frequency that is at least total-mass
dependent as in \eqref{eq:fISCO}. Such additional complications do not spoil
the accuracy of the estimate, as the following simple calculation shows. Assume
the waveforms of two systems $h_1$, $h_2$, with total masses $M_1 < M_2$. Their
distance is based on an integral in Fourier space, and due to $f_{\max}^{(1)} >
f_{\max}^{(2)} $ we can simply expand
\begin{equation}
\| h_1 - h_2 \|^2 = \Big [ \| h_1 - h_2 \|^2 \Big]_{f_{\min}}^{f_{\max}^{(2)}}
+ \Big [ \| h_1 \|^2 \Big]_{f_{\max}^{(2)}}^{f_{\max}^{(1)}},
\end{equation}
where the brackets indicate the integration range, and the second part only
contains $h_1$ because $h_2$ has been terminated already in this frequency
range. The first part can accurately be estimated with a parameter-independent
Fisher matrix that incorporates the smaller upper cutoff frequency. The second
part is proportional to a simple integral over $f^{-7/3}/S_n(f)$ in the
respective frequency range. Both contributions have been included in the data
shown in Fig.~\ref{fig:compFisherMethods}.

As evident from Fig.~\ref{fig:compFisherMethods}, the true confidence interval
is considerably smaller than predicted by ``standard'' Fisher-matrix estimates.
From efficiently calculating waveform differences for many neighboring points,
we can now simply read off the range of physical parameters that fulfills
\eqref{eq:90percInterval}, and we find
\begin{align} \label{eq:parameter_errors_real}
\frac{\Delta \eta}\eta &\lesssim 50\%, & \Delta \chi_1 &\lesssim  0.25,
\nonumber
\\
\frac{\Delta M}{M} & \lesssim 40\%, & \frac{\Delta \mathcal M}{\mathcal M} &
\lesssim 0.2\%.
\end{align}
Expressed in individual masses, we find 
\begin{equation}
2.5M_\odot \leq m_1 \leq 8.0M_\odot, \quad 1.0 M_\odot \leq m_2 \leq 2.5
M_\odot.
\end{equation}
Hence, at 90\% confidence, we would not be able to tell purely from a \ac{GW}
observation whether the observed system is composed of two rather heavy and
hardly spinning \acp{NS}, or a light \ac{NS} and a significantly spinning
\ac{BH}. The same conclusion was recently drawn in a detailed study by Hannam
\emph{et al.}~\cite{Hannam:2013uu}.

\subsection{Accurately measurable principal components}
\label{sec:PCA_interpretation}

Apart from computational benefits, the method of diagonalizing the Fisher
matrix, thereby finding uncorrelated parameters, enables us to systematically
understand which combinations of physical parameters are well measurable and,
in turn, along which paths \ac{GW} signals are almost degenerate. We consider
again our example of a $5M_\odot$/$1.35M_\odot$ \ac{BH}/\ac{NS} system with a
\ac{BH} spin of 0.3. 

We start with the standard Fisher matrix in terms of the physical parameters
$\{\log M, \log \eta, \chi_1, t_0, \phi_0 \}$. Note again that the Fisher
matrix in terms of these parameters is strongly parameter dependent, and
the results we shall obtain below are only valid for the considered target
system. Nevertheless, we present them as an instructive illustration of the
basic method before moving on to a more general interpretation.

After projecting out the time and phase shift [see
Eq.~\eqref{eq:Fisher_projected}], the eigenvalues we find are separated by 4
orders of magnitude, respectively, and the first principal component (with the
highest eigenvalue) reads
\begin{equation}
\mu_1 \propto \log M + 0.59 \log \eta - 0.02 \chi_1. 
\end{equation}
The spin dependence is rather small, so we neglect it for simplicity and find
that $\tilde \mu_1 = e^{\mu_1} \sim M \eta^{0.59}$ is remarkably similar to the
chirp mass \eqref{eq:Mchirp}. It is not surprising, but reassuring that the
\ac{PCA} indeed finds the parameter that has already been considered as the
best-measured quantity as the first principal component. Note, however, that
the spread of $\mu_1$ in the 90\% confidence interval is $\Delta \mu_1 = \Delta
\tilde \mu_1 / \tilde \mu_1 = 0.007\%$, therefore much smaller than the
variation in $\mathcal M$, cf.\ \eqref{eq:MchirpError}. We can easily
understand this by imagining the long ellipsoidal shape of the estimated
interval that extends to very different lengths along the principal components.
A minimal rotation (such as from $\tilde \mu_1$ to $\mathcal M$) can
dramatically increase the extent of the ellipse along the given direction.

Nevertheless, it is important to keep in mind that the results of the above
analysis will slightly change with different values of the source parameters,
different  variants of the detector noise curve and other cutoff frequencies.
Thus, it is likely to be more practical to consider $\mathcal M$ as the
approximately best measured parameter. It is still much more accurately
determined than the second principal component $\mu_2$, so we proceed with
calculating the Fisher matrix in terms of $\{\log \mathcal M, \eta, \chi_1, t_0,
\phi_0 \}$. After projection and diagonalization, $\mathcal M$ is indeed the
dominating contribution to $\mu_1$, and $\mu_2$ reads
\begin{equation} \label{eq:mu2_physicalpar}
\mu_2 \propto \eta + 0.42 \chi_1 .
\end{equation}
We have neglected the small $\log \mathcal M$ contribution (that has an
estimated coefficient of $0.05$) in \eqref{eq:mu2_physicalpar} as we regard the
chirp mass as essentially measured by $\mu_1$ and we are now interested in
determining the remaining parameters. As empirically observed and discussed in
Sec.~\ref{sec:diffparametrizations}, we cannot simply determine the individual
masses by measuring the two ``best'' parameters very accurately. Even though
the chirp mass is only mass dependent, the next principal component is a
combination of (symmetric) mass ratio and spin. Thus, as long as the spin value
is not determined by other means, we cannot neglect it. Neglecting it 
would result in the mass ratio being measured incorrectly. 

\begin{figure}
 \includegraphics[width=0.4\textwidth]{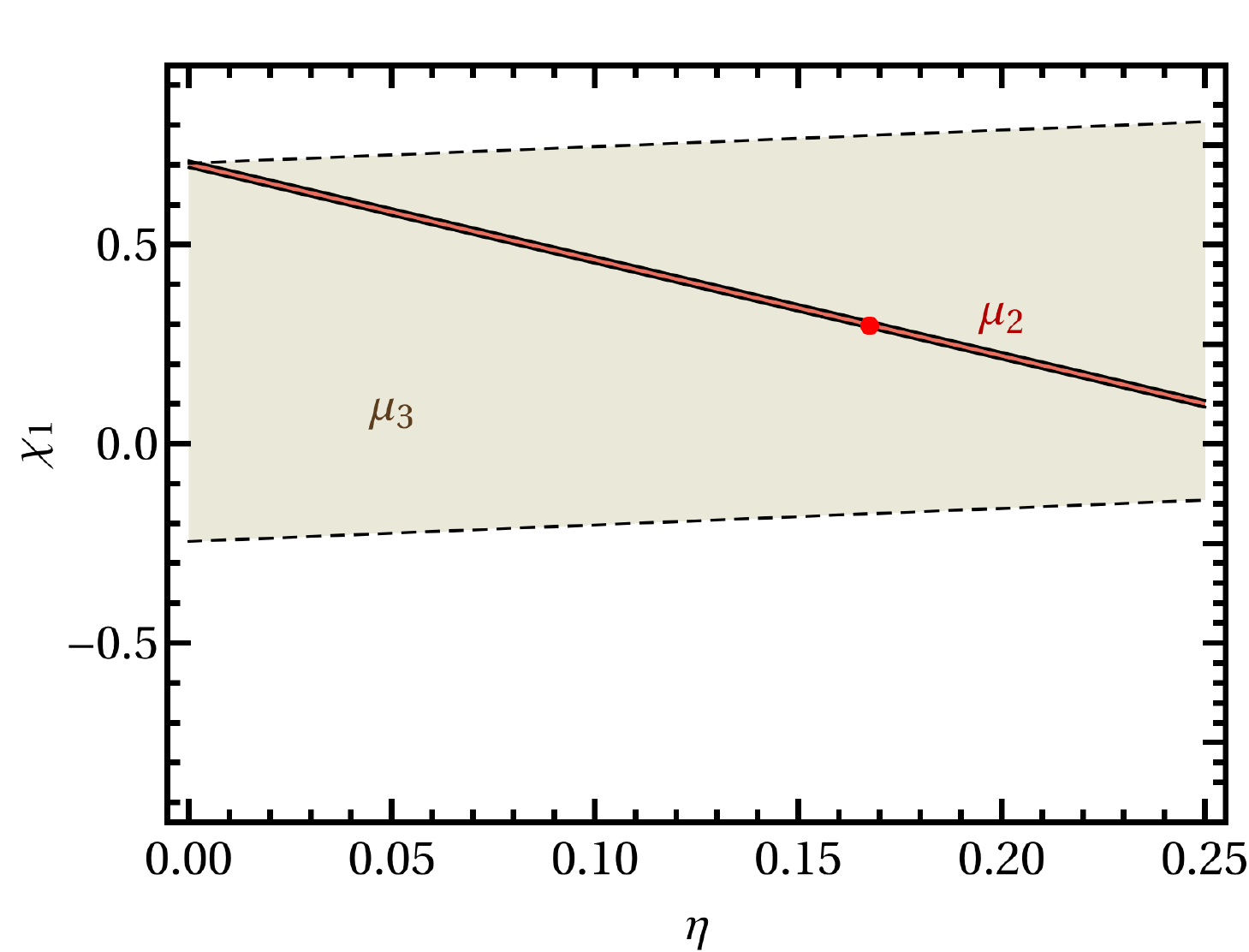}
 \caption{Illustration of the mass-ratio/spin degeneracy through principal
components for the same target signal as shown in
Fig.~\ref{fig:compFisherMethods}. Even though $\mu_2$ is accurately measurable
($\approx 1\%$), it
only restricts the parameter space to a line in the $\eta$-$\chi_1$ space. The
next principal component, $\mu_3$, is less well determined ($\approx 200\%$)
and restricts the parameter space only weakly. The best measured parameter
$\mu_1$ determines the chirp mass and does not impact this plot.}
 \label{fig:principalComp}
\end{figure}

The second principal component within the 90\% confidence interval is uncertain
by $\Delta \mu_2 / \mu_2 \approx 1 \%$, which by itself is not a large
uncertainty. However, to extract the physically more interesting parameters
$\eta$ and $\chi_1$ and with them the individual masses, we need to consider the
third principal component as well, which 
reads
\begin{equation}
\mu_3 \propto \eta - 2.40 \chi_1.
\end{equation}
We again neglect the small $\log \mathcal M$ contribution here (entering with a
factor $-0.02$). This third principal component, however, is measurable only by
$\Delta \mu_3/\mu_3 \approx 200\%$ which severely limits our ability to
identify the physical parameters. A visualization of the range of parameters
restricted by the spread in $\mu_2$ and $\mu_3$ is shown in
Fig.~\ref{fig:principalComp}. Note that this is entirely consistent with the
standard Fisher-matrix ellipse in Fig.~\ref{fig:compFisherMethods} and the
numbers presented in \eqref{eq:standardFisher_errors}. In particular, we see
that due to the tilt of $\mu_2$ in the $\eta$-$\chi_1$ plane and the fact that
$\eta$ can only take values between $0$ and $0.25$, the allowed spin is
actually somewhat restricted, whereas we cannot restrict the mass ratio of the
observed system at all, at 90\% confidence (and the assumptions underlying the
detector and waveform model).

The arguments we have just provided are simple and instructive, but, just as in
Sec.~\ref{sec:diffparametrizations}, the results are not very accurate, and the
specific functional forms of $\mu_2$ and $\mu_3$ vary throughout the parameter
space. Again, the better suited parametrization in terms of the \PN
coefficients can potentially cure both problems to some extent, because we have
already demonstrated that the Fisher-matrix estimates are much more accurate in
this higher-dimensional space. In addition to that, by using the \PN
coefficients we automatically impose a waveform-adapted functional dependence
upon the physical parameters that will lead to principal components that are
not only simple linear combinations of the source parameters.

\begin{table}
\begin{ruledtabular}
 \begin{tabular}{rrrrrrrrr}
$i$ & $\Lambda_{i1}$ & $\Lambda_{i2}$ & $\Lambda_{i3}$ & $\Lambda_{i4}$ &
$\Lambda_{i5}$ & $\Lambda_{i6}$ & $\Lambda_{i7}$ & $\Lambda_{i8}$ \\ &
\footnotesize \it 0PN &\footnotesize \it 1PN  & \footnotesize \it 1.5PN&
\footnotesize  \it 2PN& \footnotesize \it 2.5PN log& \footnotesize \it 3PN&
\footnotesize \it 3PN log& \footnotesize \it 3.5PN  \\ \hline 1&  0.98 & 0.17 &
0.06 & 0.02 & -0.03 & 0.00 & 0.00 & 0.00 \\ 2&
 -0.18 & 0.77 & 0.45 & 0.17 & -0.36 & -0.07 & -0.10 & -0.07 \\ 3&
 0.05 & -0.47 & 0.07 & 0.17 & -0.65 & -0.20 & -0.46 & -0.25 \\ 4&
 0.02 & -0.32 & 0.45 & 0.25 & -0.27 & 0.09 & 0.68 & 0.30 \\ 5&
 0.01 & -0.23 & 0.71 & 0.07 & 0.53 & 0.11 & -0.30 & -0.24 \\ 6&
 0.00 & -0.05 & 0.22 & -0.45 & -0.03 & -0.32 & -0.30 & 0.74 \\ 7& 
 0.00 & 0.02 & -0.19 & 0.77 & 0.15 & 0.15 & -0.32 & 0.47 \\ 8&
 0.00 & 0.00 & -0.04 & 0.27 & 0.25 & -0.90 & 0.20 & -0.13 
\end{tabular}
\end{ruledtabular}
\caption{Contributions of \PN phase coefficients to principal components,
obtained with a normalization frequency $f_0 = 200\,{\rm Hz}$. The cutoff
frequency employed here is $f_{\max} \approx 700\,{\rm Hz}$ (cutoff of the
reference signal in Fig.~\ref{fig:compFisherMethods}), although this affects
the highest values only weakly.}
\label{tab:principalComp_coeff_lin}
\end{table}

Let us stress again that using the \PN coefficient \eqref{eq:PNcoeffpar} as
free parameters makes the Fisher matrix only weakly varying throughout the
parameter space, thus we can analyze principal components for an entire range
of source parameters by diagonalizing just one matrix. The transformation
matrix $\Lambda_{ij}$ encodes which \PN coefficients contribute most
significantly to each principal component, which we illustrate in
Table~\ref{tab:principalComp_coeff_lin}. We have chosen the cutoff frequency to
be that of our previous target signal, i.e., \eqref{eq:fISCO} with $M =
(1.35+5) M_\odot$, $f_{\max} \approx 700\, {\rm Hz}$. However, we also tested
cutoff frequencies between $300\, {\rm Hz}$ and $2000\, {\rm Hz}$ (which is even
beyond the tidal disruption frequency for our example system, but a reasonable
cutoff for lower mass systems), and the important contributions in the first two
principal components vary by less than 10\%. 

It is worth pointing out that the numbers in
Table~\ref{tab:principalComp_coeff_lin} depend on the normalization frequency
$f_0$ chosen in \eqref{eq:PNexpancoeffs}, and it is a well-known ambiguity of
any \ac{PCA} that it changes with scale variations of the used variables. We
have chosen $f_0 = 200\,{\rm Hz}$ such that
Table~\ref{tab:principalComp_coeff_lin} gives a good indication of which \PN
coefficients are important in the Advanced LIGO detector band, but
$\Lambda_{ij}$ alone are not invariant quantities. The invariant quantity is
the waveform difference in the form of Eq.~\eqref{eq:wfdiff_eigenvec}, and our
aim is to illustrate individual contributions to this difference from various
principal components.  Therefore, by visualizing the ($f_0$-dependent) values
of $\mu_i$, we can immediately gauge how strongly the \ac{GW} signal changes
throughout the parameter space. To do that, we interpret $\mu_i$ as a function
of the physical parameters by first expressing the individual principal
components as linear combinations of \PN coefficients according to
\eqref{eq:mu_definition}. We then replace each \PN coefficient by the
appropriate phase expansion term that in turn depends on $M$, $\eta$ and
$\chi_1$.

\begin{figure}
 \includegraphics[width=0.4\textwidth]{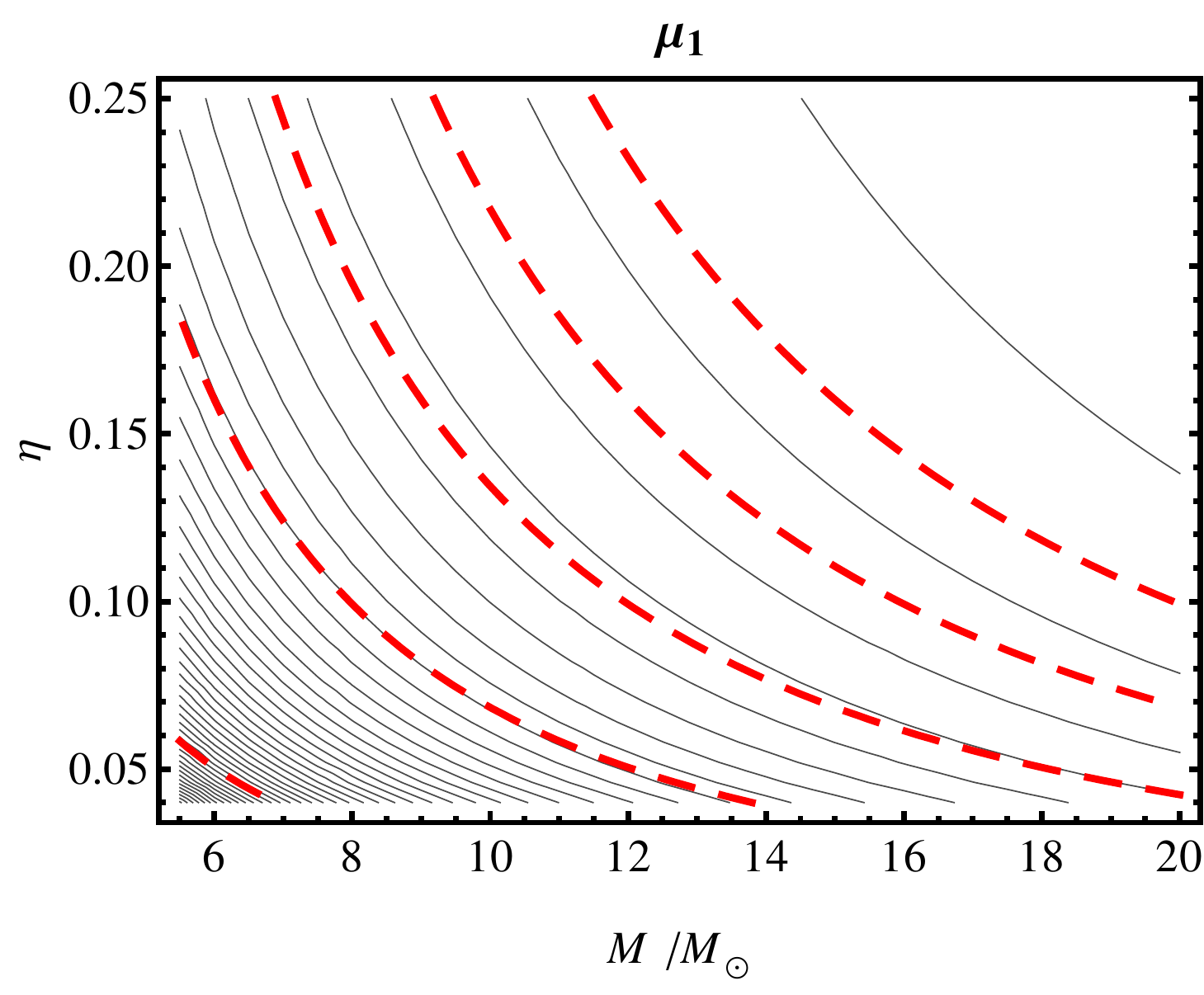}
 \caption{Contours of the first principal component $\mu_1$ (solid lines),
decreasing from bottom left to top right in steps of 1000 times the predicted
accuracy within a 90\%
confidence interval, $\Delta \mu_1^{\rm cont} = 1000 \Delta \mu_1$ (Fisher),
cf.\ Table~\ref{tab:PNcoefferrors}. Here we assume nonspinning binaries,
although the picture is largely independent of the spin. Dashed (red online)
lines are contours of constant chirp mass \eqref{eq:Mchirp}, 
$\mathcal M/M_\odot = \{1,2,3,4,5\}$ (increasing from left to right).}
 \label{fig:mu1}
\end{figure}

Figure~\ref{fig:mu1} shows contours of the constant first principal component.%
We plot contours in steps of 1000 times the predicted accuracy of $\mu_1$ in a
90\% confidence interval (see Table~\ref{tab:PNcoefferrors}), hence we see again
that $\mu_1$ is exceptionally well measurable. In addition, by simple comparison
with constant chirp-mass lines or by the fact that the Newtonian phase
contribution in $\mu_1$ is clearly dominating, we confirm once more, in a
systematic way, that the chirp mass is to a good approximation the best
measurable parameter in \ac{GW} inspiral waveforms. Furthermore, we see in
Fig.~\ref{fig:mu1} that smaller chirp masses can be measured more accurately (in
absolute terms), because the spacing between $\mu_1$ contours decreases
towards the bottom left which allows for a fine waveform distinction
perpendicular to constant-$\mathcal M$ lines.

However, we should keep in
mind that the actual best-measurable parameter is a \ac{PN}-like expansion
series with not only a $\mathcal M$-dependent dominant term, but also $\eta$-
and $\chi_1$-dependent higher-order corrections. Indeed, Fig.~\ref{fig:mu1}
does not change very sensitively with varying spin, but we find noticeable
deviations of $\mu_1$ contours from constant-$\mathcal M$ lines when the spin
parameter is close to $-1$.

\begin{figure*}
 \includegraphics[width=\textwidth]{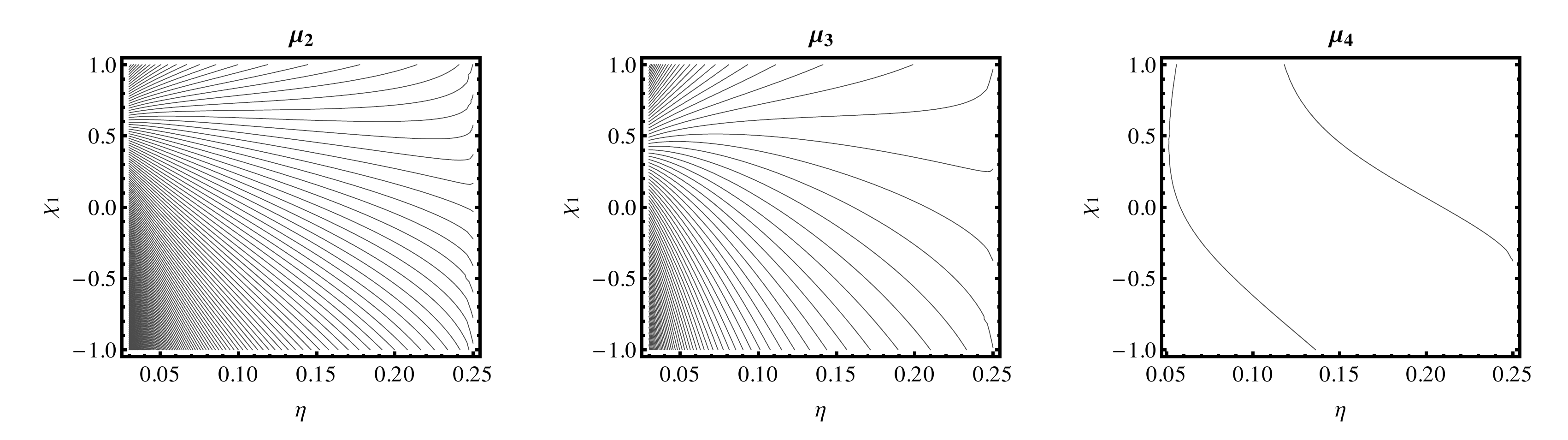}
 \caption{Contours of principal components $\mu_i$ ($i=2,3,4$), obtained from a
\ac{PCA} in terms of \PN phase coefficients, see
Table~\ref{tab:principalComp_coeff_lin}. We show slices of constant $\mu_1$
(essentially constant chirp mass).  Contours of $\mu_3$ and $\mu_4$ are drawn
in steps of $2\Delta \mu_i$ (Fisher) as given in Table~\ref{tab:PNcoefferrors},
i.e., two neighboring lines visualize the 90\% confidence interval around any
point located on the imaginary line centered within this interval. For better
readability, contours of $\mu_2$ are drawn in 5 times bigger steps, i.e., $10
\Delta \mu_2$ (Fisher). Values of the contours increase from bottom to top in
all cases.}
 \label{fig:mu2-4}
\end{figure*}

\begin{figure}
\includegraphics[width=0.4\textwidth]{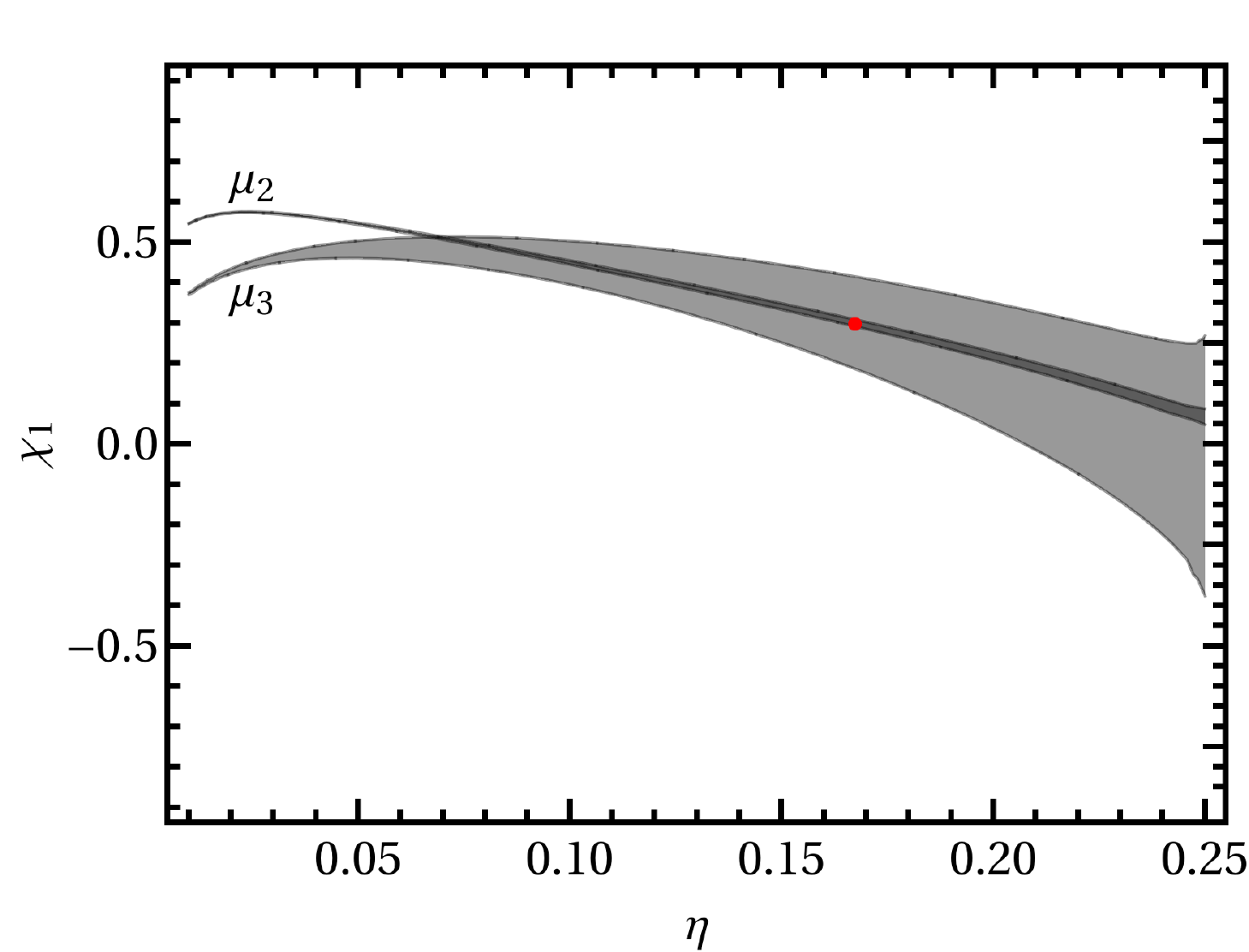}
 \caption{Combination of two principal component ranges from
Fig.~\ref{fig:mu2-4} around the target signal indicated by a (red online)
circle. The intersection of both ranges is a good approximation of the 90\%
confidence interval around the target parameters (cf.\ the actual interval in
Fig.~\ref{fig:compFisherMethods}), and it is more accurate than the estimate in
Fig.~\ref{fig:principalComp}. It is still only an upper bound because the
waveform difference is actually governed by the square sum of deviations in
$\mu_i$ \eqref{eq:wfdiff_eigenvec} and we neglect the effect of a
mass-dependent cutoff frequency here.}
 \label{fig:muConfInterval}
\end{figure}

In any case, $\mu_1$ can be very well constrained by \ac{GW} measurements, and
we use this fact to analyze the other principal components in the
$\eta$-$\chi_1$ plane only.  The other physical parameter, the total mass $M$,
is then determined for each point by inverting $\mu_1 (M, \eta, \chi_1) = {\rm
const}$, where we use the value of $\mu_1$ that corresponds to our target
signal (see Fig.~\ref{fig:compFisherMethods}) as the constant on the right-hand
side. Figure~\ref{fig:mu2-4} illustrates the resulting contours of $\mu_i$, $i=
2,3,4$. We find that both $\mu_2$ and $\mu_3$ are reasonably well measurable,
i.e., after detecting a signal, we cannot only be confident about the value of
the chirp mass (under the simplifying assumptions made here), the associated
values of $\mu_2$ and $\mu_3$ also restrict the range of plausible source
parameters to rather narrow bands in the mass-ratio/spin space. However, these
two bands have very similar structure, and accurately identifying the values of
$\eta$ and $\chi_1$ individually remains hard. This issue is illustrated in
Fig.~\ref{fig:muConfInterval}, where we overlay the predicted confidence
intervals of $\mu_2$ and $\mu_3$ around our fiducial target signal, and the
result we find is entirely consistent with the confidence interval depicted in
Fig.~\ref{fig:compFisherMethods}. Note that adding information from higher
principal components $\mu_i$, $i \geq 4$, does not add any more constraints as
their uncertainty is too large, which can be seen for $\mu_4$ in the right
panel of Fig.~\ref{fig:mu2-4}. In fact, as mentioned earlier, in the
specific case we consider the waveforms only depend on three physical
parameters, hence a fourth principal component such as $\mu_4$ cannot add any
information for determining physical parameters.

However, it is important to realize that the conclusion that three parameters
are measurable is largely independent of the functional form of the \ac{PN}
phase coefficients. One might be tempted to relate this fact to the three
physical parameters we started with, but we just showed that three
well-constrained principal components do not automatically imply that the
physical parameters can be determined to the same accuracy. Even more important
is the inverse conclusion. If we had a waveform model in the form
\eqref{eq:PNexpancoeffs}, but with phase coefficients that are functions of more
parameters (e.g., a second spin or tidal parameters of the \NS), we would only
be able to constrain three parameter combinations ($\mu_1$, $\mu_2$, $\mu_3$),
except when the variation of the \PN coefficients $\psi_k^{(\log)}$ through the
parameter space of interest is dramatically increased. Of course, the functional
form of the principal components might be different, thus
Figs.~\ref{fig:mu1}-\ref{fig:muConfInterval} may change, but1 the \ac{PCA} is
performed before the phase coefficients have to be specified, so
Table~\ref{tab:principalComp_coeff_lin} and the eigenvalues in
Table~\ref{tab:PNcoefferrors} are generally valid.

For convenience and future reference, it is useful to explicitly write down a
simplified version of $\mu_2$ that includes all spin contributions that are
known for the relevant terms. According to
Table~\ref{tab:principalComp_coeff_lin}, all 8 \PN phase coefficients enter
$\mu_2$ with nonvanishing contributions, some of which, however, are much
smaller than others. We compared the full $\mu_2$ expansion with simplified
versions that included only the one (two, three) most dominant $\Lambda_{2j}$
contribution(s). The values of such simplified expressions are different to the
full $\mu_2$ result, but since we are interested in variations of $\mu_2$
rather than the values itself, we only have to make sure that the structure of
a simplified $\mu_2$ preserves the one shown in the left panel of
Fig.~\ref{fig:mu2-4}. (The same argument is underlying our identification of
$\mathcal M$ as the approximately best-measurable parameter.) We find including
only the 1PN and 1.5PN phase terms is not sufficient, but by also adding the
2.5PN log term, we find reasonable agreement between full and simplified
$\mu_2$. 

We therefore conclude
\begin{equation}
\mu_2 \approx 0.77 \psi_2 + 0.45 \psi_3 - 0.36 \psi_5^{\log}, 
\label{eq:mu2PN}
\end{equation}
where all \PN coefficients are understood as the phase contribution at $f_0 =
200\,{\rm Hz}$. Note again that results of \acp{PCA} generally depend on the
initial choice of parameters, and there is no fundamental principle which would
guarantee that \eqref{eq:mu2PN} is similar to $\mu_2$ found earlier in terms of
physical parameters, \eqref{eq:mu2_physicalpar}. However, here we find indeed
that a linear expansion of \eqref{eq:mu2PN} in $\eta$ and $\chi_1$ for constant
$\mu_1$ yields \eqref{eq:mu2_physicalpar}. Put differently, the linear tangent
to the constant-$\mu_2$ line at the point of our target signal in
Fig.~\ref{fig:mu2-4} is accurately described by \eqref{eq:mu2_physicalpar}.
While this consistency is reassuring for assigning some physical meaning to the
principal components, we point out it does not hold for $\mu_3$ or any higher
components that only exist in the unphysical eight-dimensional space.

For convenience of the reader, we explicitly detail the phase coefficients
appearing in \eqref{eq:mu2PN} below for the more general case of two spinning
bodies, with spins aligned to the angular momentum (recall, the \ac{PCA} remains
unaffected if the \ac{NS} would be spinning, too). In the form used in
\cite{Brown:2007jx,FrankThesis}, the \PN phase coefficients read 
\begin{align}
\psi_2 &= \frac{1}{\pi M f_0} \left( \frac{55}{384} + \frac{3715}{32256 \eta}
\right),  \label{eq:psi2} \\
\psi_3 &= \frac{1}{(\pi  M f_0)^{2/3}} \left[ \frac{113}{128 \eta} \left(
\chi_s + \delta \chi_a \right) - \frac{3 \pi }{8 \eta }-\frac{19 \chi_s}{32}
\right], \label{eq:psi3}
\end{align}
%
%
\begin{align} \label{eq:psi5log}
\psi_5^{\rm log}  &=  \frac{38645 \pi }{32256 \eta } -\frac{65 \pi}{384}
-\chi_s \left(\frac{735505}{96768 \eta } - \frac{12265}{1728} - \frac{85 \eta
}{96} \right) \nonumber \\
& - \delta \chi_a \left( \frac{735505}{96768 \eta }+\frac{65}{192} \right) +
\chi_s \frac{5  (3 \eta -1)}{64 \eta }  \left( 3 \chi_a^2 + \chi_s^2 \right)
\nonumber \\
& + \delta \chi_a \frac{5  (\eta -1 )}{64 \eta}  \left( 3 \chi_s^2 + \chi_a^2
\right) ,
\end{align} 
where we used $\delta = (m_1 - m_2)/M$ and the spin combinations
\begin{align}
\chi_s = \frac{\chi_1 + \chi_2}{2}, \qquad \chi_a = \frac{\chi_1 - \chi_2}{2}.
\end{align}

It is interesting to note that while $\psi_2$ is spin independent, $\psi_3$ and
$\psi_5^{\log}$ contain the leading order and next-to-leading-order spin-orbit
terms, respectively \cite{Blanchet:2006gy}. The terms cubic in the spins are
due to the energy flow into the \acp{BH} \cite{Alvi:2001mx}. These are less
important and not valid for \acp{NS}. However, quadratic self-spin terms
\cite{Mikoczi:2005dn} and quadrupole-monopole contributions
\cite{Poisson:1997ha} that appear at relative 2\PN order (i.e., they are part
of $\psi_4$) affect the overall signal less strongly, as they have no
significant contribution to the second principal component $\mu_2$. The same
applies to even higher spin-orbit terms at 3\PN order \cite{Blanchet:2011zv}.
We shall verify the importance of particular \ac{PN} spin contributions in the
next section properly, but the results are already indicated by the form of the
principal components. 

We refrain from analyzing $\mu_3$ in the same detail. It is mainly determined
by $\psi_5^{\log}$, $\psi_2$ and $\psi_6^{\log}$ and is thereby sensitive to
even higher spin corrections. Also the highest order we consider, 3.5\PN
($\psi_7$), influences $\mu_3$ to considerably larger extent than $\mu_1$ or
$\mu_2$. We thus expect that, of the first three principal components, $\mu_3$
will be the most sensitive to higher, as yet uncalculated, \PN coefficients.
This may imply that the detectors are indeed sensitive to changes in the values
of higher order corrections to the PN expansion, even if their absolute value
is small relative to lower order terms. However, once more \PN terms have been
calculated, they can easily be included in our algorithm and the waveform model
can be analyzed for degeneracies with the method presented here.

\section{Model dependence and parameter biases -- Systematic error}
\label{Sec:biases}

While the previous section focused on the confusion caused by very similar
waveforms within \emph{one} (perfectly known) waveform manifold, we shall now
turn to systematic errors in GW measurements caused by the \emph{imperfect}
knowledge of the waveform family itself. Put differently, we shall estimate in
this section how the recovered source parameters and signal strengths are
affected when a given signal (the target signal) is not necessarily part of the
waveform manifold that is employed in the search. 

Fortunately, as long as both the target and search waveforms can be expressed
in the form of \eqref{eq:h_definition} and \eqref{eq:PNexpancoeffs}, we can
still use the linear Fisher-matrix approximation in terms of the PN expansion
coefficients $\theta_i$, Eq.~\eqref{eq:PNcoeffpar}, to estimate the effect of
different waveform families. The only difference to Sec.~\ref{sec:staterr} is
that now the \PN phase coefficients change not only due to a change of the
physical source parameters, but also due to a different functional form
$\psi_k^{(\log)} =  \psi_k^{(\log)}(M, \eta, \chi_1)$. This means that the
following study will be restricted to TaylorF2-like waveform representations;
however, we are free to modify or even drop individual PN contributions to
quantify their importance in a way meaningful for data-analysis applications.

Our strategy is as follows. Just as outlined in Sec.~\ref{sec:PCA_intro}, we
use the Fisher matrix \eqref{eq:Fisher} in terms of the \PN phase coefficients
\eqref{eq:PNcoeffpar}, and the transformation to principal components detailed
in Table~\ref{tab:principalComp_coeff_lin} is valid independently of the
functional form of the parameters. Thus, we pick a target signal by fixing the
source parameters and reference model and transform to the principal components
as usual,
\begin{equation} \label{eq:muhat}
\hat \mu_i = \sum_j \Lambda_{ij} \; \hat \theta_j .
\end{equation}
We then consider a different search model and transform from the associated \PN
parameters to the same space of principal components, such that
\begin{equation}
 \Delta \mu_i = \mu_i - \hat \mu_i = \sum_j \Lambda_{ij} \; (\theta_j - \hat
\theta_j).
\end{equation}
Again, there is no difference to what we did to analyze statistical errors,
just now $\theta_j$ will be different from $\hat \theta_j$ for the same set of
physical parameters. We can, however, vary the parameters of the search model
to minimize the difference
\begin{equation} \label{eq:minDeltah}
\min_{M, \eta, \chi_1} \| \Delta h \|^2 = \min_{M, \eta, \chi_1} \sum_i
\lambda_i \, (\Delta \mu_i)^2 . 
\end{equation}
Note that we effectively also minimize over time and phase shifts, but this is
implicit in our method through the projection of the associated parameters. The
difference between target parameters and those that minimize
\eqref{eq:minDeltah} are referred to as \emph{systematic biases}, and they
indicate to what extent a model-based \ac{GW} search would be confused by the
use of an incomplete waveform model.

As an illustration, let us consider the following scenario. The target signal
we assume is that for aligned-spin binaries including all known spin
corrections as detailed in Sec.~\ref{sec:wvmodels}.  This is the waveform model
we analyzed in Sec.~\ref{sec:staterr}. Fixing the masses again to $m_1 = 5
M_\odot$, $m_2 = 1.35 M_\odot$, we now ask the question: \emph{How well can the
mass parameters be recovered if the \ac{BH} is possibly spinning and the
employed search waveform model is that for nonspinning objects?} We easily
address this question by using standard minimization techniques (we employ a
grid-based minimization followed by a local minimization along the gradient) and
find the values that minimize \eqref{eq:minDeltah}.  Recall, the target
parameters define $\hat \mu_i$ and the variable $\mu_i$ are closed-form
functions of $M$ and $\eta$ (or equivalent parametrizations). As discussed
above, we do not need to employ sophisticated template bank generation
algorithms nor calculate direct overlaps between any waveforms to answer this
question.

\begin{figure}
 \includegraphics[width=0.42\textwidth]{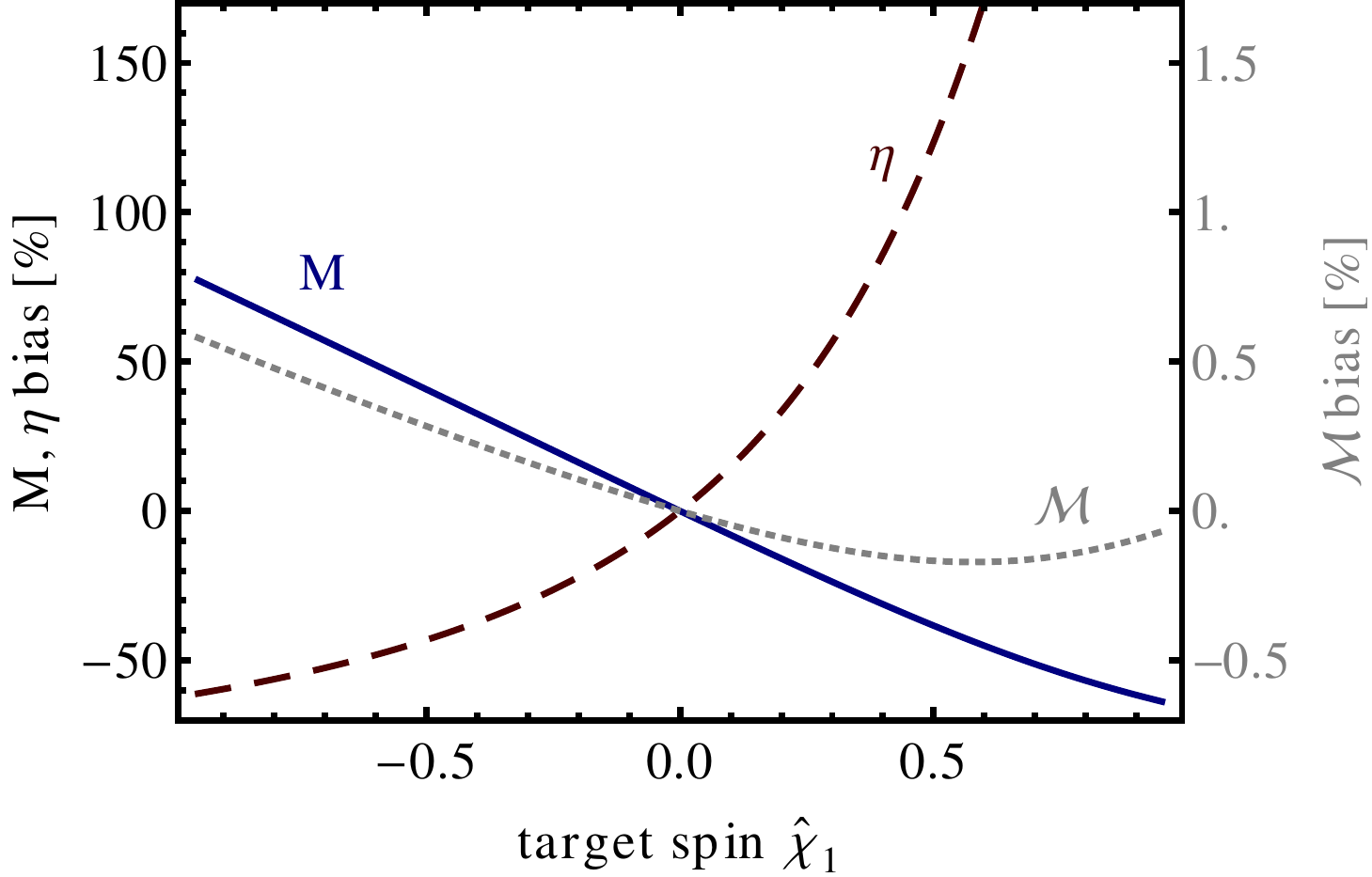}
\caption{Systematic bias introduced by a nonspinning model searching for the
waveform of a binary with a $1.35M_\odot$ nonspinning NS and a $5M_\odot$ BH
with aligned spin as indicated by the horizontal axis. The solid (blue online)
line shows the bias in the recovered total mass, the dashed (red online) line
indicates the bias in the symmetric mass ratio \eqref{eq:symm_massratio_M}. The
gray dotted line shows the bias in the chirp mass \eqref{eq:Mchirp}, for which
the scale on the right-hand side is valid.}
\label{fig:spin_vs_nonsp}
\end{figure}

The result of this exercise is shown in Fig.~\ref{fig:spin_vs_nonsp}. Not
surprisingly, we find that the bias in the chirp mass $\mathcal M$ is rather
small, which is due to the fact that the first principal component is dominated
by this quantity. The second mass parameter, either total mass $M$ or symmetric
mass ratio $\eta$, however, can be massively biased if the search does not
include spin when the source is characterized by a considerable spin. We
did not restrict the best-matching parameters to physically meaningful ranges,
and for target spins $\hat \chi_1 \gtrsim 0.3$ we find that $\eta$ exceeds its
physical range beyond $0.25$. We could have anticipated this already from the
confidence interval shown in Fig.~\ref{fig:compFisherMethods} (and further
interpreted in Figs.~\ref{fig:principalComp} and \ref{fig:muConfInterval}),
because there the 90\% confidence interval, extrapolated by eye, would
intersect the $\chi_1 = 0$ line at larger, unphysical values of $\eta$.

Apart from the bias in the recovered parameters, the actual agreement between
the best-match waveform and the target signal is of interest, as this
constitutes an estimate of the detection efficiency (i.e., how many signals
would be lost in the search due to an imperfect agreement between signal and
search family). As many authors in the \ac{GW} literature before, we shall
express the effectualness of the waveform family by the fitting factor
\begin{equation} \label{eq:FF}
{\rm FF} = \max_{\theta_{\rm ph}} \frac{\left \langle h(\theta),  h (\hat
\theta) \right \rangle}{\| h(\theta) \| \| h(\hat \theta) \|} \approx 1 -
\min_{\theta_{\rm ph}} \frac{\| h(\theta) - h(\hat \theta) \|^2}{2 \| h \|^2} ,
\end{equation}
where we optimize over the entire waveform family, i.e., over all physical
parameters $\theta_{\rm ph}$ (we added the subscript to distinguish the actual
freedom in the  waveform manifold from the higher dimensional parametrization
we are using in this paper). The right-hand side of \eqref{eq:FF} can be
calculated efficiently from the outcome of \eqref{eq:minDeltah}. This equals
the fitting factor under the assumptions that $\| h(\theta ) \| = \| h(\hat
\theta) \| = \| h \|$ \cite{Flanagan:1997kp,McWilliams:2010eq,Ohme:2011rm},
which is true for our simplified model when we neglect the variable cutoff
frequency. 

The fitting factor that corresponds to Fig.~\ref{fig:spin_vs_nonsp} deviates
from unity by as much as 15\% for highly negative \ac{BH} spins and less than
2\% for positive spin values if we allow for unphysical symmetric mass ratios.
If we restrict $\eta \leq 0.25$ then the fitting factor drops without bound for
increasing spin to an extent where we cannot trust our approximation of the
inner product any more. 

Comparing spinning against nonspinning models was a rather extreme case for
illustration, and we shall now turn to smaller differences between the target
model (which we keep fixed as the best model that includes all known spin
terms) and the search family. We are particularly interested in the effect of
various spin contributions to the \PN phasing, which we will successively drop
from the search model to analyze how well this ``reduced'' model can identify
the original signal. 

We restrict our study to the case we considered before of a $1.35M_\odot$
\ac{NS} and a $5M_\odot$ \ac{BH}, but we simulate all \ac{BH} spins $-0.95 \leq
\chi_1 \leq 0.95$ in steps of $0.05$. For each of these target signals we
minimize the difference \eqref{eq:minDeltah} to various search models and
record the fitting factor as well as the parameter biases. Each entry of
Table~\ref{tab:bias} corresponds to the simulated spin value that was
recovered with the maximal disagreement in terms of fitting factor and bias,
respectively.

\begin{table}%
\begin{ruledtabular}
 \begin{tabular}{cllll}
Model & $1-{\rm FF} ~[\%]$ & $\frac{\vert\Delta \mathcal M \vert}{\mathcal M} ~[\%]$ &  $\frac{\vert \Delta \eta \vert}{\eta} ~[\%]$ & $\vert \Delta \chi_1 \vert$ \\ \hline
no SO tail & 0.44 & 0.20 &  46 & 0.24  \\
3.5/2.5 & 0.25 & 0.06 & 28 & 0.51  \emph{(0.23)} \\
3.5/2.0 & 24 \emph{(810)} & 0.23 \emph{(0.48)} & 49 & 1.31 \emph{(0.63)} \\
no $\chi_1^2$ & 0.13 & 0.08 & 35 & 0.32 \emph{(0.20)} \\
 3.5/1.5 & 13 \emph{(450)} & 0.18 \emph{(0.22)} & 49 & 1.04 \emph{(0.63)} \\
 2.5/2.5 & 0.61 \emph{(1.02)} & 0.50 & 74 & 1.54 \emph{(0.60)}
 \end{tabular}
\end{ruledtabular}
\caption{Systematic errors of various models (see text) searching for the
waveforms that include all known \PN terms. The simulated target signals are
from $5M_\odot+1.35 M_\odot$ binaries with the heavier object spinning with
$-0.95 \leq \hat \chi_1 \leq 0.95$.  We report the maximal disagreement between
the best-fit model waveforms and the target signals in terms of the fitting
factor \eqref{eq:FF} and the biases in chirp mass, symmetric mass ratio and
spin. Searches were performed with unrestricted and restricted spin range, and
italic numbers in parentheses indicate the values in case we obtain
significantly different numbers when the search space is restricted to
physically meaningful spins $-1 \leq \chi_1 \leq 1$.}
\label{tab:bias}
\end{table}

The search models we consider are as follows:
\begin{description}[font=\normalfont,itemsep=0pt]
 \item[\it no SO tail] Up to 3.5PN nonspinning and spinning contributions
included (incomplete 3PN and 3.5PN spin corrections inherited from re-expanded
lower-order terms, see discussion in \cite{Hannam:2010ec}), but without the 3PN
spin-orbit tail contribution derived in \cite{Blanchet:2011zv};
 \item[\it 3.5/2.5PN] Up to 3.5PN nonspinning and up to 2.5PN spinning
contributions included, i.e., no incomplete spin terms considered;
 \item[\it 3.5/2.0PN] Up to 3.5PN nonspinning and up to 2.0PN spinning
contributions included, i.e., next-to-leading-order spin-orbit coupling
dropped;
 \item[\it no $\chi_1^2$] Same as 3.5/2.5PN, but without quadratic spin terms
at 2PN order;
 \item[\it 3.5/1.5] Up to 3.5PN nonspinning contributions included plus only
the leading order spin-orbit coupling at 1.5PN;
 \item[\it 2.5/2.5] Up to 2.5PN spinning and nonspinning contributions
included.
\end{description}
Interestingly, we find from the results in Table~\ref{tab:bias} that the
reduced search models have reasonably high fitting factor with the full target
waveform if at least the dominant \emph{and} next-to-leading order spin-orbit
coupling are included in the model. Higher spin-orbit contributions, quadratic
self-spin terms, quadrupole-monopole interactions and even higher-order
nonspinning corrections are less important for the detection of the signal. 

The systematic biases are not as easily interpretable, because every search
model exhibits almost degenerate regions of parameter space with
indistinguishable waveforms, similar to what we have analyzed in
Sec.~\ref{sec:staterr}. Thus, a template waveform with a much lower or higher
parameter bias might agree with the target signal almost equally well, but the
point we report as the result of a numerical optimization procedure does not
include this information. We can, however, compare the results in
Table~\ref{tab:bias} with the statistical uncertainty reported under the
assumption of a perfectly known waveform family,
Eq.~\eqref{eq:parameter_errors_real}, which leads us to the conclusion
that the systematic errors reported in Table~\ref{tab:bias} are 
acceptable for the models with low fitting factor, except the 2.5/2.5PN case
where nonspinning contributions are truncated.

It is important, however, to point out that neglecting the 2.5PN spin-orbit
coupling leads to a severe loss in $\rm FF$, and searches employing only the
leading-order spin corrections are prone to miss signals from binaries with
considerable spin. The numbers in Table~\ref{tab:bias} were obtained by
allowing unbounded values for the spin of the waveform model, and indeed,
particularly the cases with low fitting factor achieve the best agreement with
unphysical values of $\chi_1 < -1$. If we instead restrict the search parameter
space to physically meaningful ranges $\vert \chi_1 \vert \leq 1$, we obtain
different numbers in some cases, given in parentheses in Table~\ref{tab:bias}.
Note that the already badly performing models then become completely
disconnected to the target waveform space which results in absurdly high
deviation of $\rm FF$ from unity. These numbers are an artifact of using the
Fisher matrix to estimate large waveform mismatches and cannot be trusted.
However, the fact that we would be unable to detect some spinning systems with
such models is only emphasized by these results.

It is interesting to note that we already observed a similar effect with
nonspinning searches and unphysical values of $\eta$, as discussed in
connection with Fig.~\ref{fig:spin_vs_nonsp}. Here we cannot allow for
unphysical $\eta$ as some spinning contributions contain $\delta = (m_1 -
m_2)/M = \sqrt{1  -4 \eta}$, see Eqs.~\eqref{eq:psi2}-\eqref{eq:psi5log}, which
has no real solution for $\eta > 0.25$. However, we have just
illustrated that unphysical values of the spin(s) may potentially inflate the
waveform manifold enough to increase the detection efficiency such that signals
that are not part of the search family have a higher chance of
being detected.

\section{Conclusions} \label{sec:conclusion}

In this paper, we have considered nonprecessing inspiral waveforms of \acp{GW}
emitted by coalescing \ac{NS}-BH binaries. Such models are essential
ingredients for the ongoing efforts to directly detect \acp{GW} for the first
time, and the success and astrophysical output of such detections will depend
sensitively on our understanding of the waveform family employed in the search.

By combining the well-known Fisher-matrix approach with a suitable
higher-dimensional coordinate choice, we have demonstrated that the analysis of
degeneracies in the waveform space can be made considerably more accurate than
previous Fisher-matrix studies of parameter measurabilities, while still much
faster than full overlap calculations between individual waveforms. Even though
the high-dimensional Fisher matrix may be ill conditioned, we argued that the
relevant information about the waveforms can be extracted through, instead of
inversion, diagonalization of the Fisher matrix. This is because only the
eigenvectors with large eigenvalues affect the waveform considerably. Thus,
this procedure (which we identify as a \PCA) is still well conditioned, and we
explicitly presented how we can efficiently find confidence intervals around a
given signal including a parameter-dependent cutoff frequency.

The coordinate choice we employed is based on assuming the \PN phase expansion
coefficients are free parameters~\cite{Tanaka:2000xy, Sathyaprakash:2003ua,
Pai:2012mv, Brown:2012qf}. This approach relies on the waveform model being
written as a simple amplitude and a complex phase which is a sum of purely
frequency-dependent functions, each multiplied by a single parameter at most.
Extending this strategy to accommodate more complicated functions, such as full
inspiral-merger-ringdown models or precessing systems, is difficult as these
models do not obey this simple analytic form. However, we restricted ourselves
to a regime where the merger and ringdown part of the signal do not contribute
significantly, and recent investigations show that modeling precessing systems
may be based on a modulation of nonprecessing signals \cite{Schmidt:2010it,
Boyle:2011gg, Schmidt:2012rh}. In addition, waveform families that are used for
detection purposes are unlikely to model full precession \cite{Ajith:2011ec,
Brown:2012gs, Ajith:2012mn}. Thus, even though realistic signals are expected
to contain some amount of precession, it is worth analyzing nonprecessing
signals first.

In agreement with previous publications \cite{Cutler:1994ys, Poisson:1995ef,
Baird:2012cu, Hannam:2013uu}, we found that the individual masses of the
binary's constituents cannot be well constrained by \GW observations alone.
This is because even though the chirp mass is measurable very accurately, the
second mass parameter can be confused by the presence of spin.  Disentangling
spin and mass ratio would require yet another measurement, which is not
accurate enough to place useful bounds on the individual masses.

With the analysis carried out in this paper, we can now rephrase these results
in a more formal manner, following the results of our \PCA. The first, very
accurately determined principal component is dominated by the chirp mass (with
higher-order spin-dependent corrections). The second principal component can be
seen as a combination of symmetric mass ratio and spin that may somewhat
restrict the spin magnitude, but does not allow for an unambiguous measurement
of either parameter. A third principal component is also measurable to
reasonable accuracy, but it adds little restrictions to the range in mass ratio
and spin in our case. Higher principal components cannot be well constrained by
\GW measurements and they do not vary enough through the parameter space of
interest to add much information. 

It is important to point out that the explicit form of the principal components
is model and gauge dependent (in our case, the scale freedom is expressed by
the normalization frequency $f_0$), so the interpretation of the waveform
structure in terms of principal components reveals no fundamental property of
the waveform manifold. It is nevertheless a useful concept to understand the
prospects and limitations of modeled \GW searches. For instance, we have
demonstrated that three parameters can be measured accurately, but whether or
not these lead to astrophysically meaningful statements has to be determined by
the explicit dependence of these well-measured parameters on physically
interesting quantities. 

This explicit form of principal directions in parameter space is, in turn,
derived in terms of \PN expansion coefficients. In our analysis, we have found
that higher-order terms also have a noticeable influence on the third principal
component, suggesting that yet undetermined nonspinning and especially spinning
corrections may influence our ability to measure parameters in the future.
Among the currently determined \PN contributions we identified the leading and
next-to-leading order spin-orbit terms as crucial spin corrections that need to
be included in the waveform model to not change the manifold drastically.
Again, our fitting-factor study of systematic errors was limited to
Fourier-domain models of the form detailed above. It would be interesting to
contrast our results with comparisons between various approximants in the time
and frequency domain. Note however, that time-domain models (such as the
TaylorT\emph{n} approximants) can, in principle, be transformed to an analytic
form in Fourier space as well, where the difference between those models and
the TaylorF2 model employed here would lie entirely in undetermined
higher-order phase corrections. Their effect can be studied in our framework by
allowing the parameter space to be extending beyond 4\PN order, which we leave
for future work. 

As long as such higher-order terms have not been fully calculated, we
need to ensure that the waveform model chosen for use in \GW searches is
capable of detecting signals from other, equally plausible models as well. One
way of doing this is by allowing unphysical source parameters. We have,
somewhat artificially, compared the full waveform model here with reduced
search families that were lacking certain contributions, which we take as a
guideline to the situation we are actually facing. Namely, that we search for
signals in the universe (that may or may not be well described by the full
theory of general relativity) with a restricted, incomplete, \PN model.  We
have demonstrated that allowing an unphysical spin parameter beyond unit
magnitude can indeed reduce the systematic difference of waveform families. It
remains to be tested more thoroughly whether, for the detection problem, that
reduces ambiguities to a negligible extent. Parameter estimation pipelines, on
the other hand, obviously cannot use this freedom. However, our algorithm also
provides an easy way to estimate parameter biases between different
waveform models with physically meaningful bounds.

In summary, our results provide a formal answer to the question of what can be
measured by \GW observations of inspiraling \NS-\BH binaries. 
This is important astrophysically, but also has some immediate applications for
standard \GW data-analysis techniques. For instance, constructing a
discrete template bank for spinning signals with predefined maximal mismatch
between templates becomes much simpler in our adapted coordinates (as detailed
in \cite{Brown:2012qf,Harry:2013tca}), and our results promote a
physical understanding of the resulting parameter-space coverage. A related
question is important for the relatively small number of numerical-relativity
waveforms that have to be calculated in order to calibrate complete
inspiral-merger-ringdown models (see \cite{Ohme:2011rm} for an overview). Their
parameter-space coverage should take advantage of the dominant directions in
waveform space that can be estimated with our method. Equally, testing
inspiral degeneracies in the merger regime is an important task
\cite{Purrer:2013ojf} that, however, requires identifying these
degeneracies first. 

In addition, our principal coordinates should allow for a geometric, very
efficient parameter estimation beyond the template with highest overlap, and
even more advanced and accurate parameter-estimation routines (such as Markov
chain Monte-Carlo methods) may benefit from knowing the preferred directions in
parameter space that we identify.

Finally, calculating inner products between different waveforms is the core of
all matched-filter algorithms, and the computationally very efficient
approximation we suggest should allow for a tremendous speed-up of all analyses
that are centered in a regime of high overlaps. We have presented an example 
study of systematic errors that otherwise had been computationally
very challenging. Now, however, such studies can quickly be 
repeated and extended for other setups (e.g., different PN coefficients,
detector noise curves, etc.) which facilitates easy sanity checks and
on-line comparisons of results obtained with different waveform families.

\section*{ACKNOWLEDGEMENTS}

It is a pleasure to thank Stephen Fairhurst, Mark Hannam, Ian Harry, Badri
Krishnan, Chris Messenger and Bangalore Sathyaprakash for many useful
discussions. We are also grateful to Francesco Pannarale for sharing his
insights into \NS spin expectations and observations, and for carefully reading
the initial manuscript. FO would also like to thank Lukas Ohme, simply for
being there when this paper became public. FO is supported by STFC Grant No.\
ST/I001085/1.

\bibliographystyle{apsrev4-1}

\bibliography{chirptimes}

\begin{thebibliography}{64}%
\makeatletter
\providecommand \@ifxundefined [1]{%
 \@ifx{#1\undefined}
}%
\providecommand \@ifnum [1]{%
 \ifnum #1\expandafter \@firstoftwo
 \else \expandafter \@secondoftwo
 \fi
}%
\providecommand \@ifx [1]{%
 \ifx #1\expandafter \@firstoftwo
 \else \expandafter \@secondoftwo
 \fi
}%
\providecommand \natexlab [1]{#1}%
\providecommand \enquote  [1]{``#1''}%
\providecommand \bibnamefont  [1]{#1}%
\providecommand \bibfnamefont [1]{#1}%
\providecommand \citenamefont [1]{#1}%
\providecommand \href@noop [0]{\@secondoftwo}%
\providecommand \href [0]{\begingroup \@sanitize@url \@href}%
\providecommand \@href[1]{\@@startlink{#1}\@@href}%
\providecommand \@@href[1]{\endgroup#1\@@endlink}%
\providecommand \@sanitize@url [0]{\catcode `\\12\catcode `\$12\catcode
  `\&12\catcode `\#12\catcode `\^12\catcode `\_12\catcode `\%12\relax}%
\providecommand \@@startlink[1]{}%
\providecommand \@@endlink[0]{}%
\providecommand \url  [0]{\begingroup\@sanitize@url \@url }%
\providecommand \@url [1]{\endgroup\@href {#1}{\urlprefix }}%
\providecommand \urlprefix  [0]{URL }%
\providecommand \Eprint [0]{\href }%
\providecommand \doibase [0]{http://dx.doi.org/}%
\providecommand \selectlanguage [0]{\@gobble}%
\providecommand \bibinfo  [0]{\@secondoftwo}%
\providecommand \bibfield  [0]{\@secondoftwo}%
\providecommand \translation [1]{[#1]}%
\providecommand \BibitemOpen [0]{}%
\providecommand \bibitemStop [0]{}%
\providecommand \bibitemNoStop [0]{.\EOS\space}%
\providecommand \EOS [0]{\spacefactor3000\relax}%
\providecommand \BibitemShut  [1]{\csname bibitem#1\endcsname}%
\let\auto@bib@innerbib\@empty
\bibitem [{\citenamefont {Abbott}\ \emph {et~al.}(2009)\citenamefont {Abbott}
  \emph {et~al.}}]{Abbott:2007kv}%
  \BibitemOpen
  \bibfield  {author} {\bibinfo {author} {\bibfnamefont {B.}~\bibnamefont
  {Abbott}} \emph {et~al.} (\bibinfo {collaboration} {LIGO Scientific
  Collaboration}),\ }\href {\doibase 10.1088/0034-4885/72/7/076901} {\bibfield
  {journal} {\bibinfo  {journal} {Rept.Prog.Phys.}\ }\textbf {\bibinfo {volume}
  {72}},\ \bibinfo {pages} {076901} (\bibinfo {year} {2009})},\ \Eprint
  {http://arxiv.org/abs/0711.3041} {arXiv:0711.3041 [gr-qc]} \BibitemShut
  {NoStop}%
\bibitem [{\citenamefont {Sigg}(2008)}]{Sigg:2008zz}%
  \BibitemOpen
  \bibfield  {author} {\bibinfo {author} {\bibfnamefont {D.}~\bibnamefont
  {Sigg}} (\bibinfo {collaboration} {LIGO Scientific Collaboration}),\ }\href
  {\doibase 10.1088/0264-9381/25/11/114041} {\bibfield  {journal} {\bibinfo
  {journal} {Class.Quant.Grav.}\ }\textbf {\bibinfo {volume} {25}},\ \bibinfo
  {pages} {114041} (\bibinfo {year} {2008})}\BibitemShut {NoStop}%
\bibitem [{\citenamefont {Smith}(2009)}]{Smith:2009bx}%
  \BibitemOpen
  \bibfield  {author} {\bibinfo {author} {\bibfnamefont {J.~R.}\ \bibnamefont
  {Smith}} (\bibinfo {collaboration} {LIGO Scientific Collaboration}),\ }\href
  {\doibase 10.1088/0264-9381/26/11/114013} {\bibfield  {journal} {\bibinfo
  {journal} {Class.Quant.Grav.}\ }\textbf {\bibinfo {volume} {26}},\ \bibinfo
  {pages} {114013} (\bibinfo {year} {2009})},\ \Eprint
  {http://arxiv.org/abs/0902.0381} {arXiv:0902.0381 [gr-qc]} \BibitemShut
  {NoStop}%
\bibitem [{\citenamefont {Harry}(2010)}]{Harry:2010zz}%
  \BibitemOpen
  \bibfield  {author} {\bibinfo {author} {\bibfnamefont {G.~M.}\ \bibnamefont
  {Harry}} (\bibinfo {collaboration} {LIGO Scientific Collaboration}),\ }\href
  {\doibase 10.1088/0264-9381/27/8/084006} {\bibfield  {journal} {\bibinfo
  {journal} {Class.Quant.Grav.}\ }\textbf {\bibinfo {volume} {27}},\ \bibinfo
  {pages} {084006} (\bibinfo {year} {2010})}\BibitemShut {NoStop}%
\bibitem [{\citenamefont {Acernese}\ \emph {et~al.}(2008)\citenamefont
  {Acernese}, \citenamefont {Alshourbagy}, \citenamefont {Amico}, \citenamefont
  {Antonucci}, \citenamefont {Aoudia} \emph {et~al.}}]{Acernese:2008zzf}%
  \BibitemOpen
  \bibfield  {author} {\bibinfo {author} {\bibfnamefont {F.}~\bibnamefont
  {Acernese}}, \bibinfo {author} {\bibfnamefont {M.}~\bibnamefont
  {Alshourbagy}}, \bibinfo {author} {\bibfnamefont {P.}~\bibnamefont {Amico}},
  \bibinfo {author} {\bibfnamefont {F.}~\bibnamefont {Antonucci}}, \bibinfo
  {author} {\bibfnamefont {S.}~\bibnamefont {Aoudia}},  \emph {et~al.},\ }\href
  {\doibase 10.1088/0264-9381/25/18/184001} {\bibfield  {journal} {\bibinfo
  {journal} {Class.Quant.Grav.}\ }\textbf {\bibinfo {volume} {25}},\ \bibinfo
  {pages} {184001} (\bibinfo {year} {2008})}\BibitemShut {NoStop}%
\bibitem [{\citenamefont {Accadia}\ \emph {et~al.}(2011)\citenamefont
  {Accadia}, \citenamefont {Acernese}, \citenamefont {Astone}, \citenamefont
  {Ballardin}, \citenamefont {Barone} \emph {et~al.}}]{Accadia:2011zz}%
  \BibitemOpen
  \bibfield  {author} {\bibinfo {author} {\bibfnamefont {T.}~\bibnamefont
  {Accadia}}, \bibinfo {author} {\bibfnamefont {F.}~\bibnamefont {Acernese}},
  \bibinfo {author} {\bibfnamefont {P.}~\bibnamefont {Astone}}, \bibinfo
  {author} {\bibfnamefont {G.}~\bibnamefont {Ballardin}}, \bibinfo {author}
  {\bibfnamefont {F.}~\bibnamefont {Barone}},  \emph {et~al.},\ }\href
  {\doibase 10.1063/1.3637466} {\bibfield  {journal} {\bibinfo  {journal}
  {Rev.Sci.Instrum.}\ }\textbf {\bibinfo {volume} {82}},\ \bibinfo {pages}
  {094502} (\bibinfo {year} {2011})}\BibitemShut {NoStop}%
\bibitem [{\citenamefont {Abadie}\ \emph {et~al.}(2010)\citenamefont {Abadie}
  \emph {et~al.}}]{Abadie:2010cf}%
  \BibitemOpen
  \bibfield  {author} {\bibinfo {author} {\bibfnamefont {J.}~\bibnamefont
  {Abadie}} \emph {et~al.} (\bibinfo {collaboration} {LIGO Scientific
  Collaboration, Virgo Collaboration}),\ }\href {\doibase
  10.1088/0264-9381/27/17/173001} {\bibfield  {journal} {\bibinfo  {journal}
  {Class.Quant.Grav.}\ }\textbf {\bibinfo {volume} {27}},\ \bibinfo {pages}
  {173001} (\bibinfo {year} {2010})},\ \Eprint {http://arxiv.org/abs/1003.2480}
  {arXiv:1003.2480 [astro-ph.HE]} \BibitemShut {NoStop}%
\bibitem [{\citenamefont {Blanchet}(2006)}]{lrr-2006-4}%
  \BibitemOpen
  \bibfield  {author} {\bibinfo {author} {\bibfnamefont {L.}~\bibnamefont
  {Blanchet}},\ }\href@noop {} {\bibfield  {journal} {\bibinfo  {journal}
  {Living Reviews in Relativity}\ }\textbf {\bibinfo {volume} {9}} (\bibinfo
  {year} {2006})},\ \bibinfo {note}
  {\url{http://www.livingreviews.org/lrr-2006-4}}\BibitemShut {NoStop}%
\bibitem [{\citenamefont {Buonanno}\ \emph {et~al.}(2009)\citenamefont
  {Buonanno}, \citenamefont {Iyer}, \citenamefont {Ochsner}, \citenamefont
  {Pan},\ and\ \citenamefont {Sathyaprakash}}]{Buonanno:2009zt}%
  \BibitemOpen
  \bibfield  {author} {\bibinfo {author} {\bibfnamefont {A.}~\bibnamefont
  {Buonanno}}, \bibinfo {author} {\bibfnamefont {B.}~\bibnamefont {Iyer}},
  \bibinfo {author} {\bibfnamefont {E.}~\bibnamefont {Ochsner}}, \bibinfo
  {author} {\bibfnamefont {Y.}~\bibnamefont {Pan}}, \ and\ \bibinfo {author}
  {\bibfnamefont {B.}~\bibnamefont {Sathyaprakash}},\ }\href {\doibase
  10.1103/PhysRevD.80.084043} {\bibfield  {journal} {\bibinfo  {journal}
  {Phys.Rev.}\ }\textbf {\bibinfo {volume} {D80}},\ \bibinfo {pages} {084043}
  (\bibinfo {year} {2009})},\ \Eprint {http://arxiv.org/abs/0907.0700}
  {arXiv:0907.0700 [gr-qc]} \BibitemShut {NoStop}%
\bibitem [{\citenamefont {Baird}\ \emph {et~al.}(2013)\citenamefont {Baird},
  \citenamefont {Fairhurst}, \citenamefont {Hannam},\ and\ \citenamefont
  {Murphy}}]{Baird:2012cu}%
  \BibitemOpen
  \bibfield  {author} {\bibinfo {author} {\bibfnamefont {E.}~\bibnamefont
  {Baird}}, \bibinfo {author} {\bibfnamefont {S.}~\bibnamefont {Fairhurst}},
  \bibinfo {author} {\bibfnamefont {M.}~\bibnamefont {Hannam}}, \ and\ \bibinfo
  {author} {\bibfnamefont {P.}~\bibnamefont {Murphy}},\ }\href {\doibase
  10.1103/PhysRevD.87.024035} {\bibfield  {journal} {\bibinfo  {journal}
  {Phys.Rev.}\ }\textbf {\bibinfo {volume} {D87}},\ \bibinfo {pages} {024035}
  (\bibinfo {year} {2013})},\ \Eprint {http://arxiv.org/abs/1211.0546}
  {arXiv:1211.0546 [gr-qc]} \BibitemShut {NoStop}%
\bibitem [{\citenamefont {Hannam}\ \emph {et~al.}(2013)\citenamefont {Hannam},
  \citenamefont {Brown}, \citenamefont {Fairhurst}, \citenamefont {Fryer},\
  and\ \citenamefont {Harry}}]{Hannam:2013uu}%
  \BibitemOpen
  \bibfield  {author} {\bibinfo {author} {\bibfnamefont {M.}~\bibnamefont
  {Hannam}}, \bibinfo {author} {\bibfnamefont {D.~A.}\ \bibnamefont {Brown}},
  \bibinfo {author} {\bibfnamefont {S.}~\bibnamefont {Fairhurst}}, \bibinfo
  {author} {\bibfnamefont {C.~L.}\ \bibnamefont {Fryer}}, \ and\ \bibinfo
  {author} {\bibfnamefont {I.~W.}\ \bibnamefont {Harry}},\ }\href {\doibase
  10.1088/2041-8205/766/1/L14} {\bibfield  {journal} {\bibinfo  {journal}
  {Astrophys.J.}\ }\textbf {\bibinfo {volume} {766}},\ \bibinfo {pages} {L14}
  (\bibinfo {year} {2013})},\ \Eprint {http://arxiv.org/abs/1301.5616}
  {arXiv:1301.5616 [gr-qc]} \BibitemShut {NoStop}%
\bibitem [{\citenamefont {Bildsten}\ and\ \citenamefont
  {Cutler}(1992)}]{Bildsten:1992my}%
  \BibitemOpen
  \bibfield  {author} {\bibinfo {author} {\bibfnamefont {L.}~\bibnamefont
  {Bildsten}}\ and\ \bibinfo {author} {\bibfnamefont {C.}~\bibnamefont
  {Cutler}},\ }\href {\doibase 10.1086/171983} {\bibfield  {journal} {\bibinfo
  {journal} {Astrophys.J.}\ }\textbf {\bibinfo {volume} {400}},\ \bibinfo
  {pages} {175} (\bibinfo {year} {1992})}\BibitemShut {NoStop}%
\bibitem [{\citenamefont {Kochanek}(1992)}]{Kochanek:1992wk}%
  \BibitemOpen
  \bibfield  {author} {\bibinfo {author} {\bibfnamefont {C.~S.}\ \bibnamefont
  {Kochanek}},\ }\href {\doibase 10.1086/171851} {\bibfield  {journal}
  {\bibinfo  {journal} {Astrophys.J.}\ }\textbf {\bibinfo {volume} {398}},\
  \bibinfo {pages} {234} (\bibinfo {year} {1992})}\BibitemShut {NoStop}%
\bibitem [{\citenamefont {Vallisneri}(2008)}]{Vallisneri:2007ev}%
  \BibitemOpen
  \bibfield  {author} {\bibinfo {author} {\bibfnamefont {M.}~\bibnamefont
  {Vallisneri}},\ }\href {\doibase 10.1103/PhysRevD.77.042001} {\bibfield
  {journal} {\bibinfo  {journal} {Phys.Rev.}\ }\textbf {\bibinfo {volume}
  {D77}},\ \bibinfo {pages} {042001} (\bibinfo {year} {2008})},\ \Eprint
  {http://arxiv.org/abs/gr-qc/0703086} {arXiv:gr-qc/0703086 [gr-qc]}
  \BibitemShut {NoStop}%
\bibitem [{\citenamefont {Tanaka}\ and\ \citenamefont
  {Tagoshi}(2000)}]{Tanaka:2000xy}%
  \BibitemOpen
  \bibfield  {author} {\bibinfo {author} {\bibfnamefont {T.}~\bibnamefont
  {Tanaka}}\ and\ \bibinfo {author} {\bibfnamefont {H.}~\bibnamefont
  {Tagoshi}},\ }\href {\doibase 10.1103/PhysRevD.62.082001} {\bibfield
  {journal} {\bibinfo  {journal} {Phys.Rev.}\ }\textbf {\bibinfo {volume}
  {D62}},\ \bibinfo {pages} {082001} (\bibinfo {year} {2000})},\ \Eprint
  {http://arxiv.org/abs/gr-qc/0001090} {arXiv:gr-qc/0001090 [gr-qc]}
  \BibitemShut {NoStop}%
\bibitem [{\citenamefont {Sathyaprakash}\ and\ \citenamefont
  {Schutz}(2003)}]{Sathyaprakash:2003ua}%
  \BibitemOpen
  \bibfield  {author} {\bibinfo {author} {\bibfnamefont {B.}~\bibnamefont
  {Sathyaprakash}}\ and\ \bibinfo {author} {\bibfnamefont {B.~F.}\ \bibnamefont
  {Schutz}},\ }\href@noop {} {\bibfield  {journal} {\bibinfo  {journal}
  {Class.Quant.Grav.}\ }\textbf {\bibinfo {volume} {20}},\ \bibinfo {pages}
  {S209} (\bibinfo {year} {2003})},\ \Eprint
  {http://arxiv.org/abs/gr-qc/0301049} {arXiv:gr-qc/0301049 [gr-qc]}
  \BibitemShut {NoStop}%
\bibitem [{\citenamefont {Pai}\ and\ \citenamefont {Arun}(2013)}]{Pai:2012mv}%
  \BibitemOpen
  \bibfield  {author} {\bibinfo {author} {\bibfnamefont {A.}~\bibnamefont
  {Pai}}\ and\ \bibinfo {author} {\bibfnamefont {K.}~\bibnamefont {Arun}},\
  }\href {\doibase 10.1088/0264-9381/30/2/025011} {\bibfield  {journal}
  {\bibinfo  {journal} {Class.Quant.Grav.}\ }\textbf {\bibinfo {volume} {30}},\
  \bibinfo {pages} {025011} (\bibinfo {year} {2013})},\ \Eprint
  {http://arxiv.org/abs/1207.1943} {arXiv:1207.1943 [gr-qc]} \BibitemShut
  {NoStop}%
\bibitem [{\citenamefont {Brown}\ \emph
  {et~al.}(2012{\natexlab{a}})\citenamefont {Brown}, \citenamefont {Harry},
  \citenamefont {Lundgren},\ and\ \citenamefont {Nitz}}]{Brown:2012qf}%
  \BibitemOpen
  \bibfield  {author} {\bibinfo {author} {\bibfnamefont {D.~A.}\ \bibnamefont
  {Brown}}, \bibinfo {author} {\bibfnamefont {I.}~\bibnamefont {Harry}},
  \bibinfo {author} {\bibfnamefont {A.}~\bibnamefont {Lundgren}}, \ and\
  \bibinfo {author} {\bibfnamefont {A.~H.}\ \bibnamefont {Nitz}},\ }\href
  {\doibase 10.1103/PhysRevD.86.084017} {\bibfield  {journal} {\bibinfo
  {journal} {Phys.Rev.}\ }\textbf {\bibinfo {volume} {D86}},\ \bibinfo {pages}
  {084017} (\bibinfo {year} {2012}{\natexlab{a}})},\ \Eprint
  {http://arxiv.org/abs/1207.6406} {arXiv:1207.6406 [gr-qc]} \BibitemShut
  {NoStop}%
\bibitem [{\citenamefont {Shoemaker}()}]{advLIGO}%
  \BibitemOpen
  \bibfield  {author} {\bibinfo {author} {\bibfnamefont {D.}~\bibnamefont
  {Shoemaker}} (\bibinfo {collaboration} {LIGO Scientific Collaboration}),\
  }\href@noop {} {\enquote {\bibinfo {title} {{Advanced LIGO anticipated
  sensitivity curves}},}\ }\bibinfo {note} {\url{https://dcc.ligo.org/cgi-bin/
  DocDB/ShowDocument?docid=2974}}\BibitemShut {NoStop}%
\bibitem [{\citenamefont {Finn}\ and\ \citenamefont
  {Chernoff}(1993)}]{Finn:1992xs}%
  \BibitemOpen
  \bibfield  {author} {\bibinfo {author} {\bibfnamefont {L.~S.}\ \bibnamefont
  {Finn}}\ and\ \bibinfo {author} {\bibfnamefont {D.~F.}\ \bibnamefont
  {Chernoff}},\ }\href {\doibase 10.1103/PhysRevD.47.2198} {\bibfield
  {journal} {\bibinfo  {journal} {Phys.Rev.}\ }\textbf {\bibinfo {volume}
  {D47}},\ \bibinfo {pages} {2198} (\bibinfo {year} {1993})},\ \Eprint
  {http://arxiv.org/abs/gr-qc/9301003} {arXiv:gr-qc/9301003 [gr-qc]}
  \BibitemShut {NoStop}%
\bibitem [{\citenamefont {Jaranowski}\ and\ \citenamefont
  {Kr{\'o}lak}(1994)}]{Jaranowski:1994xd}%
  \BibitemOpen
  \bibfield  {author} {\bibinfo {author} {\bibfnamefont {P.}~\bibnamefont
  {Jaranowski}}\ and\ \bibinfo {author} {\bibfnamefont {A.}~\bibnamefont
  {Kr{\'o}lak}},\ }\href {\doibase 10.1103/PhysRevD.49.1723} {\bibfield
  {journal} {\bibinfo  {journal} {Phys.Rev.}\ }\textbf {\bibinfo {volume}
  {D49}},\ \bibinfo {pages} {1723} (\bibinfo {year} {1994})}\BibitemShut
  {NoStop}%
\bibitem [{\citenamefont {Cutler}\ and\ \citenamefont
  {Flanagan}(1994)}]{Cutler:1994ys}%
  \BibitemOpen
  \bibfield  {author} {\bibinfo {author} {\bibfnamefont {C.}~\bibnamefont
  {Cutler}}\ and\ \bibinfo {author} {\bibfnamefont {E.~E.}\ \bibnamefont
  {Flanagan}},\ }\href {\doibase 10.1103/PhysRevD.49.2658} {\bibfield
  {journal} {\bibinfo  {journal} {Phys.Rev.}\ }\textbf {\bibinfo {volume}
  {D49}},\ \bibinfo {pages} {2658} (\bibinfo {year} {1994})},\ \Eprint
  {http://arxiv.org/abs/gr-qc/9402014} {arXiv:gr-qc/9402014 [gr-qc]}
  \BibitemShut {NoStop}%
\bibitem [{\citenamefont {Poisson}\ and\ \citenamefont
  {Will}(1995)}]{Poisson:1995ef}%
  \BibitemOpen
  \bibfield  {author} {\bibinfo {author} {\bibfnamefont {E.}~\bibnamefont
  {Poisson}}\ and\ \bibinfo {author} {\bibfnamefont {C.~M.}\ \bibnamefont
  {Will}},\ }\href {\doibase 10.1103/PhysRevD.52.848} {\bibfield  {journal}
  {\bibinfo  {journal} {Phys.Rev.}\ }\textbf {\bibinfo {volume} {D52}},\
  \bibinfo {pages} {848} (\bibinfo {year} {1995})},\ \Eprint
  {http://arxiv.org/abs/gr-qc/9502040} {arXiv:gr-qc/9502040 [gr-qc]}
  \BibitemShut {NoStop}%
\bibitem [{\citenamefont {Nielsen}(2013)}]{Nielsen:2012sb}%
  \BibitemOpen
  \bibfield  {author} {\bibinfo {author} {\bibfnamefont {A.~B.}\ \bibnamefont
  {Nielsen}},\ }\href@noop {} {\bibfield  {journal} {\bibinfo  {journal}
  {Class.Quant.Grav.}\ }\textbf {\bibinfo {volume} {30}},\ \bibinfo {pages}
  {075023} (\bibinfo {year} {2013})},\ \Eprint {http://arxiv.org/abs/1203.6603}
  {arXiv:1203.6603 [gr-qc]} \BibitemShut {NoStop}%
\bibitem [{\citenamefont {Arun}\ \emph {et~al.}(2005)\citenamefont {Arun},
  \citenamefont {Iyer}, \citenamefont {Sathyaprakash},\ and\ \citenamefont
  {Sundararajan}}]{Arun:2004hn}%
  \BibitemOpen
  \bibfield  {author} {\bibinfo {author} {\bibfnamefont {K.}~\bibnamefont
  {Arun}}, \bibinfo {author} {\bibfnamefont {B.~R.}\ \bibnamefont {Iyer}},
  \bibinfo {author} {\bibfnamefont {B.}~\bibnamefont {Sathyaprakash}}, \ and\
  \bibinfo {author} {\bibfnamefont {P.~A.}\ \bibnamefont {Sundararajan}},\
  }\href {\doibase 10.1103/PhysRevD.71.084008, 10.1103/PhysRevD.72.069903}
  {\bibfield  {journal} {\bibinfo  {journal} {Phys.Rev.}\ }\textbf {\bibinfo
  {volume} {D71}},\ \bibinfo {pages} {084008} (\bibinfo {year} {2005})},\
  \Eprint {http://arxiv.org/abs/gr-qc/0411146} {arXiv:gr-qc/0411146 [gr-qc]}
  \BibitemShut {NoStop}%
\bibitem [{\citenamefont {Lindblom}\ \emph {et~al.}(2008)\citenamefont
  {Lindblom}, \citenamefont {Owen},\ and\ \citenamefont
  {Brown}}]{Lindblom:2008cm}%
  \BibitemOpen
  \bibfield  {author} {\bibinfo {author} {\bibfnamefont {L.}~\bibnamefont
  {Lindblom}}, \bibinfo {author} {\bibfnamefont {B.~J.}\ \bibnamefont {Owen}},
  \ and\ \bibinfo {author} {\bibfnamefont {D.~A.}\ \bibnamefont {Brown}},\
  }\href {\doibase 10.1103/PhysRevD.78.124020} {\bibfield  {journal} {\bibinfo
  {journal} {Phys. Rev.}\ }\textbf {\bibinfo {volume} {D78}},\ \bibinfo {pages}
  {124020} (\bibinfo {year} {2008})},\ \Eprint {http://arxiv.org/abs/0809.3844}
  {arXiv:0809.3844 [gr-qc]} \BibitemShut {NoStop}%
\bibitem [{\citenamefont {Lindblom}(2009)}]{Lindblom:2009ux}%
  \BibitemOpen
  \bibfield  {author} {\bibinfo {author} {\bibfnamefont {L.}~\bibnamefont
  {Lindblom}},\ }\href {\doibase 10.1103/PhysRevD.80.064019} {\bibfield
  {journal} {\bibinfo  {journal} {Phys.Rev.}\ }\textbf {\bibinfo {volume}
  {D80}},\ \bibinfo {pages} {064019} (\bibinfo {year} {2009})},\ \Eprint
  {http://arxiv.org/abs/0907.0457} {arXiv:0907.0457 [gr-qc]} \BibitemShut
  {NoStop}%
\bibitem [{\citenamefont {Damour}\ \emph {et~al.}(2011)\citenamefont {Damour},
  \citenamefont {Nagar},\ and\ \citenamefont {Trias}}]{Damour:2010zb}%
  \BibitemOpen
  \bibfield  {author} {\bibinfo {author} {\bibfnamefont {T.}~\bibnamefont
  {Damour}}, \bibinfo {author} {\bibfnamefont {A.}~\bibnamefont {Nagar}}, \
  and\ \bibinfo {author} {\bibfnamefont {M.}~\bibnamefont {Trias}},\ }\href
  {\doibase 10.1103/PhysRevD.83.024006} {\bibfield  {journal} {\bibinfo
  {journal} {Phys.Rev.}\ }\textbf {\bibinfo {volume} {D83}},\ \bibinfo {pages}
  {024006} (\bibinfo {year} {2011})},\ \Eprint {http://arxiv.org/abs/1009.5998}
  {arXiv:1009.5998 [gr-qc]} \BibitemShut {NoStop}%
\bibitem [{\citenamefont {Damour}\ \emph {et~al.}(2001)\citenamefont {Damour},
  \citenamefont {Iyer},\ and\ \citenamefont {Sathyaprakash}}]{Damour:2000zb}%
  \BibitemOpen
  \bibfield  {author} {\bibinfo {author} {\bibfnamefont {T.}~\bibnamefont
  {Damour}}, \bibinfo {author} {\bibfnamefont {B.~R.}\ \bibnamefont {Iyer}}, \
  and\ \bibinfo {author} {\bibfnamefont {B.}~\bibnamefont {Sathyaprakash}},\
  }\href {\doibase 10.1103/PhysRevD.63.044023, 10.1103/PhysRevD.72.029902}
  {\bibfield  {journal} {\bibinfo  {journal} {Phys.Rev.}\ }\textbf {\bibinfo
  {volume} {D63}},\ \bibinfo {pages} {044023} (\bibinfo {year} {2001})},\
  \Eprint {http://arxiv.org/abs/gr-qc/0010009} {arXiv:gr-qc/0010009 [gr-qc]}
  \BibitemShut {NoStop}%
\bibitem [{\citenamefont {Damour}\ \emph {et~al.}(2002)\citenamefont {Damour},
  \citenamefont {Iyer},\ and\ \citenamefont {Sathyaprakash}}]{Damour:2002kr}%
  \BibitemOpen
  \bibfield  {author} {\bibinfo {author} {\bibfnamefont {T.}~\bibnamefont
  {Damour}}, \bibinfo {author} {\bibfnamefont {B.~R.}\ \bibnamefont {Iyer}}, \
  and\ \bibinfo {author} {\bibfnamefont {B.~S.}\ \bibnamefont
  {Sathyaprakash}},\ }\href {\doibase 10.1103/PhysRevD.66.027502} {\bibfield
  {journal} {\bibinfo  {journal} {Phys. Rev.}\ }\textbf {\bibinfo {volume}
  {D66}},\ \bibinfo {pages} {027502} (\bibinfo {year} {2002})},\ \Eprint
  {http://arxiv.org/abs/gr-qc/0207021} {arXiv:gr-qc/0207021} \BibitemShut
  {NoStop}%
\bibitem [{\citenamefont {Blanchet}\ \emph {et~al.}(2008)\citenamefont
  {Blanchet}, \citenamefont {Faye}, \citenamefont {Iyer},\ and\ \citenamefont
  {Sinha}}]{Blanchet:2008je}%
  \BibitemOpen
  \bibfield  {author} {\bibinfo {author} {\bibfnamefont {L.}~\bibnamefont
  {Blanchet}}, \bibinfo {author} {\bibfnamefont {G.}~\bibnamefont {Faye}},
  \bibinfo {author} {\bibfnamefont {B.~R.}\ \bibnamefont {Iyer}}, \ and\
  \bibinfo {author} {\bibfnamefont {S.}~\bibnamefont {Sinha}},\ }\href@noop {}
  {\bibfield  {journal} {\bibinfo  {journal} {Class.Quant.Grav.}\ }\textbf
  {\bibinfo {volume} {25}},\ \bibinfo {pages} {165003} (\bibinfo {year}
  {2008})},\ \Eprint {http://arxiv.org/abs/0802.1249} {arXiv:0802.1249 [gr-qc]}
  \BibitemShut {NoStop}%
\bibitem [{\citenamefont {Arun}\ \emph {et~al.}(2009)\citenamefont {Arun},
  \citenamefont {Buonanno}, \citenamefont {Faye},\ and\ \citenamefont
  {Ochsner}}]{Arun:2008kb}%
  \BibitemOpen
  \bibfield  {author} {\bibinfo {author} {\bibfnamefont {K.~G.}\ \bibnamefont
  {Arun}}, \bibinfo {author} {\bibfnamefont {A.}~\bibnamefont {Buonanno}},
  \bibinfo {author} {\bibfnamefont {G.}~\bibnamefont {Faye}}, \ and\ \bibinfo
  {author} {\bibfnamefont {E.}~\bibnamefont {Ochsner}},\ }\href {\doibase
  10.1103/PhysRevD.79.104023} {\bibfield  {journal} {\bibinfo  {journal} {Phys.
  Rev.}\ }\textbf {\bibinfo {volume} {D79}},\ \bibinfo {pages} {104023}
  (\bibinfo {year} {2009})},\ \Eprint {http://arxiv.org/abs/0810.5336}
  {arXiv:0810.5336 [gr-qc]} \BibitemShut {NoStop}%
\bibitem [{\citenamefont {Kidder}(1995)}]{Kidder:1995zr}%
  \BibitemOpen
  \bibfield  {author} {\bibinfo {author} {\bibfnamefont {L.~E.}\ \bibnamefont
  {Kidder}},\ }\href {\doibase 10.1103/PhysRevD.52.821} {\bibfield  {journal}
  {\bibinfo  {journal} {Phys.Rev.}\ }\textbf {\bibinfo {volume} {D52}},\
  \bibinfo {pages} {821} (\bibinfo {year} {1995})},\ \Eprint
  {http://arxiv.org/abs/gr-qc/9506022} {arXiv:gr-qc/9506022 [gr-qc]}
  \BibitemShut {NoStop}%
\bibitem [{\citenamefont {Apostolatos}\ \emph {et~al.}(1994)\citenamefont
  {Apostolatos}, \citenamefont {Cutler}, \citenamefont {Sussman},\ and\
  \citenamefont {Thorne}}]{Apostolatos:1994mx}%
  \BibitemOpen
  \bibfield  {author} {\bibinfo {author} {\bibfnamefont {T.~A.}\ \bibnamefont
  {Apostolatos}}, \bibinfo {author} {\bibfnamefont {C.}~\bibnamefont {Cutler}},
  \bibinfo {author} {\bibfnamefont {G.~J.}\ \bibnamefont {Sussman}}, \ and\
  \bibinfo {author} {\bibfnamefont {K.~S.}\ \bibnamefont {Thorne}},\ }\href
  {\doibase 10.1103/PhysRevD.49.6274} {\bibfield  {journal} {\bibinfo
  {journal} {Phys.Rev.}\ }\textbf {\bibinfo {volume} {D49}},\ \bibinfo {pages}
  {6274} (\bibinfo {year} {1994})}\BibitemShut {NoStop}%
\bibitem [{\citenamefont {Blanchet}\ \emph {et~al.}(2006)\citenamefont
  {Blanchet}, \citenamefont {Buonanno},\ and\ \citenamefont
  {Faye}}]{Blanchet:2006gy}%
  \BibitemOpen
  \bibfield  {author} {\bibinfo {author} {\bibfnamefont {L.}~\bibnamefont
  {Blanchet}}, \bibinfo {author} {\bibfnamefont {A.}~\bibnamefont {Buonanno}},
  \ and\ \bibinfo {author} {\bibfnamefont {G.}~\bibnamefont {Faye}},\ }\href
  {\doibase 10.1103/PhysRevD.81.089901, 10.1103/PhysRevD.75.049903,
  10.1103/PhysRevD.74.104034} {\bibfield  {journal} {\bibinfo  {journal}
  {Phys.Rev.}\ }\textbf {\bibinfo {volume} {D74}},\ \bibinfo {pages} {104034}
  (\bibinfo {year} {2006})},\ \Eprint {http://arxiv.org/abs/gr-qc/0605140}
  {arXiv:gr-qc/0605140 [gr-qc]} \BibitemShut {NoStop}%
\bibitem [{\citenamefont {Blanchet}\ \emph {et~al.}(2011)\citenamefont
  {Blanchet}, \citenamefont {Buonanno},\ and\ \citenamefont
  {Faye}}]{Blanchet:2011zv}%
  \BibitemOpen
  \bibfield  {author} {\bibinfo {author} {\bibfnamefont {L.}~\bibnamefont
  {Blanchet}}, \bibinfo {author} {\bibfnamefont {A.}~\bibnamefont {Buonanno}},
  \ and\ \bibinfo {author} {\bibfnamefont {G.}~\bibnamefont {Faye}},\ }\href
  {\doibase 10.1103/PhysRevD.84.064041} {\bibfield  {journal} {\bibinfo
  {journal} {Phys.Rev.}\ }\textbf {\bibinfo {volume} {D84}},\ \bibinfo {pages}
  {064041} (\bibinfo {year} {2011})},\ \Eprint {http://arxiv.org/abs/1104.5659}
  {arXiv:1104.5659 [gr-qc]} \BibitemShut {NoStop}%
\bibitem [{\citenamefont {Damour}(2001)}]{Damour:2001tu}%
  \BibitemOpen
  \bibfield  {author} {\bibinfo {author} {\bibfnamefont {T.}~\bibnamefont
  {Damour}},\ }\href {\doibase 10.1103/PhysRevD.64.124013} {\bibfield
  {journal} {\bibinfo  {journal} {Phys.Rev.}\ }\textbf {\bibinfo {volume}
  {D64}},\ \bibinfo {pages} {124013} (\bibinfo {year} {2001})},\ \Eprint
  {http://arxiv.org/abs/gr-qc/0103018} {arXiv:gr-qc/0103018 [gr-qc]}
  \BibitemShut {NoStop}%
\bibitem [{\citenamefont {Poisson}(1998)}]{Poisson:1997ha}%
  \BibitemOpen
  \bibfield  {author} {\bibinfo {author} {\bibfnamefont {E.}~\bibnamefont
  {Poisson}},\ }\href {\doibase 10.1103/PhysRevD.57.5287} {\bibfield  {journal}
  {\bibinfo  {journal} {Phys.Rev.}\ }\textbf {\bibinfo {volume} {D57}},\
  \bibinfo {pages} {5287} (\bibinfo {year} {1998})},\ \Eprint
  {http://arxiv.org/abs/gr-qc/9709032} {arXiv:gr-qc/9709032 [gr-qc]}
  \BibitemShut {NoStop}%
\bibitem [{\citenamefont {Mikoczi}\ \emph {et~al.}(2005)\citenamefont
  {Mikoczi}, \citenamefont {Vasuth},\ and\ \citenamefont
  {Gergely}}]{Mikoczi:2005dn}%
  \BibitemOpen
  \bibfield  {author} {\bibinfo {author} {\bibfnamefont {B.}~\bibnamefont
  {Mikoczi}}, \bibinfo {author} {\bibfnamefont {M.}~\bibnamefont {Vasuth}}, \
  and\ \bibinfo {author} {\bibfnamefont {L.~A.}\ \bibnamefont {Gergely}},\
  }\href {\doibase 10.1103/PhysRevD.71.124043} {\bibfield  {journal} {\bibinfo
  {journal} {Phys.Rev.}\ }\textbf {\bibinfo {volume} {D71}},\ \bibinfo {pages}
  {124043} (\bibinfo {year} {2005})},\ \Eprint
  {http://arxiv.org/abs/astro-ph/0504538} {arXiv:astro-ph/0504538 [astro-ph]}
  \BibitemShut {NoStop}%
\bibitem [{\citenamefont {Alvi}(2001)}]{Alvi:2001mx}%
  \BibitemOpen
  \bibfield  {author} {\bibinfo {author} {\bibfnamefont {K.}~\bibnamefont
  {Alvi}},\ }\href {\doibase 10.1103/PhysRevD.64.104020} {\bibfield  {journal}
  {\bibinfo  {journal} {Phys. Rev.}\ }\textbf {\bibinfo {volume} {D64}},\
  \bibinfo {pages} {104020} (\bibinfo {year} {2001})},\ \Eprint
  {http://arxiv.org/abs/gr-qc/0107080} {arXiv:gr-qc/0107080} \BibitemShut
  {NoStop}%
\bibitem [{\citenamefont {Ohme}(2012{\natexlab{a}})}]{FrankThesis}%
  \BibitemOpen
  \bibfield  {author} {\bibinfo {author} {\bibfnamefont {F.}~\bibnamefont
  {Ohme}},\ }\emph {\bibinfo {title} {Bridging the gap between post-Newtonian
  theory and numerical relativity in gravitational-wave data analysis}},\
  \href@noop {} {Ph.D. thesis},\ \bibinfo  {school} {Potsdam University}
  (\bibinfo {year} {2012}{\natexlab{a}}),\ \bibinfo {note}
  {\url{http://nbn-resolving.de/urn:nbn:de:kobv:517-opus-60346}}\BibitemShut
  {NoStop}%
\bibitem [{\citenamefont {Ajith}\ \emph {et~al.}(2007)\citenamefont {Ajith}
  \emph {et~al.}}]{Brown:2007jx}%
  \BibitemOpen
  \bibfield  {author} {\bibinfo {author} {\bibfnamefont {P.}~\bibnamefont
  {Ajith}} \emph {et~al.},\ }\href@noop {} {\  (\bibinfo {year} {2007})},\
  \Eprint {http://arxiv.org/abs/0709.0093} {arXiv:0709.0093 [gr-qc]}
  \BibitemShut {NoStop}%
\bibitem [{\citenamefont {Bohé}\ \emph {et~al.}(2013)\citenamefont {Bohé},
  \citenamefont {Marsat},\ and\ \citenamefont {Blanchet}}]{Bohe:2013cla}%
  \BibitemOpen
  \bibfield  {author} {\bibinfo {author} {\bibfnamefont {A.}~\bibnamefont
  {Bohé}}, \bibinfo {author} {\bibfnamefont {S.}~\bibnamefont {Marsat}}, \
  and\ \bibinfo {author} {\bibfnamefont {L.}~\bibnamefont {Blanchet}},\ }\href
  {\doibase 10.1088/0264-9381/30/13/135009} {\bibfield  {journal} {\bibinfo
  {journal} {Class.Quant.Grav.}\ }\textbf {\bibinfo {volume} {30}},\ \bibinfo
  {pages} {135009} (\bibinfo {year} {2013})},\ \Eprint
  {http://arxiv.org/abs/1303.7412} {arXiv:1303.7412 [gr-qc]} \BibitemShut
  {NoStop}%
\bibitem [{\citenamefont {Burgay}\ \emph {et~al.}(2003)\citenamefont {Burgay},
  \citenamefont {D'Amico}, \citenamefont {Possenti}, \citenamefont
  {Manchester}, \citenamefont {Lyne} \emph {et~al.}}]{Burgay:2003jj}%
  \BibitemOpen
  \bibfield  {author} {\bibinfo {author} {\bibfnamefont {M.}~\bibnamefont
  {Burgay}}, \bibinfo {author} {\bibfnamefont {N.}~\bibnamefont {D'Amico}},
  \bibinfo {author} {\bibfnamefont {A.}~\bibnamefont {Possenti}}, \bibinfo
  {author} {\bibfnamefont {R.}~\bibnamefont {Manchester}}, \bibinfo {author}
  {\bibfnamefont {A.}~\bibnamefont {Lyne}},  \emph {et~al.},\ }\href {\doibase
  10.1038/nature02124} {\bibfield  {journal} {\bibinfo  {journal} {Nature}\
  }\textbf {\bibinfo {volume} {426}},\ \bibinfo {pages} {531} (\bibinfo {year}
  {2003})},\ \Eprint {http://arxiv.org/abs/astro-ph/0312071}
  {arXiv:astro-ph/0312071 [astro-ph]} \BibitemShut {NoStop}%
\bibitem [{\citenamefont {Ajith}\ \emph {et~al.}(2011)\citenamefont {Ajith},
  \citenamefont {Hannam}, \citenamefont {Husa}, \citenamefont {Chen},
  \citenamefont {Br{\"u}gmann} \emph {et~al.}}]{Ajith:2009bn}%
  \BibitemOpen
  \bibfield  {author} {\bibinfo {author} {\bibfnamefont {P.}~\bibnamefont
  {Ajith}}, \bibinfo {author} {\bibfnamefont {M.}~\bibnamefont {Hannam}},
  \bibinfo {author} {\bibfnamefont {S.}~\bibnamefont {Husa}}, \bibinfo {author}
  {\bibfnamefont {Y.}~\bibnamefont {Chen}}, \bibinfo {author} {\bibfnamefont
  {B.}~\bibnamefont {Br{\"u}gmann}},  \emph {et~al.},\ }\href {\doibase
  10.1103/PhysRevLett.106.241101} {\bibfield  {journal} {\bibinfo  {journal}
  {Phys.Rev.Lett.}\ }\textbf {\bibinfo {volume} {106}},\ \bibinfo {pages}
  {241101} (\bibinfo {year} {2011})},\ \Eprint {http://arxiv.org/abs/0909.2867}
  {arXiv:0909.2867 [gr-qc]} \BibitemShut {NoStop}%
\bibitem [{\citenamefont {Santamar{\'i}a}\ \emph {et~al.}(2010)\citenamefont
  {Santamar{\'i}a}, \citenamefont {Ohme}, \citenamefont {Ajith}, \citenamefont
  {Br{\"u}gmann}, \citenamefont {Dorband} \emph {et~al.}}]{Santamaria:2010yb}%
  \BibitemOpen
  \bibfield  {author} {\bibinfo {author} {\bibfnamefont {L.}~\bibnamefont
  {Santamar{\'i}a}}, \bibinfo {author} {\bibfnamefont {F.}~\bibnamefont
  {Ohme}}, \bibinfo {author} {\bibfnamefont {P.}~\bibnamefont {Ajith}},
  \bibinfo {author} {\bibfnamefont {B.}~\bibnamefont {Br{\"u}gmann}}, \bibinfo
  {author} {\bibfnamefont {N.}~\bibnamefont {Dorband}},  \emph {et~al.},\
  }\href {\doibase 10.1103/PhysRevD.82.064016} {\bibfield  {journal} {\bibinfo
  {journal} {Phys.Rev.}\ }\textbf {\bibinfo {volume} {D82}},\ \bibinfo {pages}
  {064016} (\bibinfo {year} {2010})},\ \Eprint {http://arxiv.org/abs/1005.3306}
  {arXiv:1005.3306 [gr-qc]} \BibitemShut {NoStop}%
\bibitem [{\citenamefont {Ajith}(2011)}]{Ajith:2011ec}%
  \BibitemOpen
  \bibfield  {author} {\bibinfo {author} {\bibfnamefont {P.}~\bibnamefont
  {Ajith}},\ }\href {\doibase 10.1103/PhysRevD.84.084037} {\bibfield  {journal}
  {\bibinfo  {journal} {Phys.Rev.}\ }\textbf {\bibinfo {volume} {D84}},\
  \bibinfo {pages} {084037} (\bibinfo {year} {2011})},\ \Eprint
  {http://arxiv.org/abs/1107.1267} {arXiv:1107.1267 [gr-qc]} \BibitemShut
  {NoStop}%
\bibitem [{\citenamefont {Yunes}\ and\ \citenamefont
  {Pretorius}(2009)}]{Yunes:2009ke}%
  \BibitemOpen
  \bibfield  {author} {\bibinfo {author} {\bibfnamefont {N.}~\bibnamefont
  {Yunes}}\ and\ \bibinfo {author} {\bibfnamefont {F.}~\bibnamefont
  {Pretorius}},\ }\href {\doibase 10.1103/PhysRevD.80.122003} {\bibfield
  {journal} {\bibinfo  {journal} {Phys.Rev.}\ }\textbf {\bibinfo {volume}
  {D80}},\ \bibinfo {pages} {122003} (\bibinfo {year} {2009})},\ \Eprint
  {http://arxiv.org/abs/0909.3328} {arXiv:0909.3328 [gr-qc]} \BibitemShut
  {NoStop}%
\bibitem [{\citenamefont {Li}\ \emph {et~al.}(2012)\citenamefont {Li},
  \citenamefont {Del~Pozzo}, \citenamefont {Vitale}, \citenamefont {Van
  Den~Broeck}, \citenamefont {Agathos} \emph {et~al.}}]{Li:2011cg}%
  \BibitemOpen
  \bibfield  {author} {\bibinfo {author} {\bibfnamefont {T.}~\bibnamefont
  {Li}}, \bibinfo {author} {\bibfnamefont {W.}~\bibnamefont {Del~Pozzo}},
  \bibinfo {author} {\bibfnamefont {S.}~\bibnamefont {Vitale}}, \bibinfo
  {author} {\bibfnamefont {C.}~\bibnamefont {Van Den~Broeck}}, \bibinfo
  {author} {\bibfnamefont {M.}~\bibnamefont {Agathos}},  \emph {et~al.},\
  }\href {\doibase 10.1103/PhysRevD.85.082003} {\bibfield  {journal} {\bibinfo
  {journal} {Phys.Rev.}\ }\textbf {\bibinfo {volume} {D85}},\ \bibinfo {pages}
  {082003} (\bibinfo {year} {2012})},\ \Eprint {http://arxiv.org/abs/1110.0530}
  {arXiv:1110.0530 [gr-qc]} \BibitemShut {NoStop}%
\bibitem [{\citenamefont {Babak}\ \emph {et~al.}(2006)\citenamefont {Babak},
  \citenamefont {Balasubramanian}, \citenamefont {Churches}, \citenamefont
  {Cokelaer},\ and\ \citenamefont {Sathyaprakash}}]{Babak:2006ty}%
  \BibitemOpen
  \bibfield  {author} {\bibinfo {author} {\bibfnamefont {S.}~\bibnamefont
  {Babak}}, \bibinfo {author} {\bibfnamefont {R.}~\bibnamefont
  {Balasubramanian}}, \bibinfo {author} {\bibfnamefont {D.}~\bibnamefont
  {Churches}}, \bibinfo {author} {\bibfnamefont {T.}~\bibnamefont {Cokelaer}},
  \ and\ \bibinfo {author} {\bibfnamefont {B.}~\bibnamefont {Sathyaprakash}},\
  }\href {\doibase 10.1088/0264-9381/23/18/002} {\bibfield  {journal} {\bibinfo
   {journal} {Class.Quant.Grav.}\ }\textbf {\bibinfo {volume} {23}},\ \bibinfo
  {pages} {5477} (\bibinfo {year} {2006})},\ \Eprint
  {http://arxiv.org/abs/gr-qc/0604037} {arXiv:gr-qc/0604037 [gr-qc]}
  \BibitemShut {NoStop}%
\bibitem [{\citenamefont {Owen}(1996)}]{Owen:1995tm}%
  \BibitemOpen
  \bibfield  {author} {\bibinfo {author} {\bibfnamefont {B.~J.}\ \bibnamefont
  {Owen}},\ }\href {\doibase 10.1103/PhysRevD.53.6749} {\bibfield  {journal}
  {\bibinfo  {journal} {Phys.Rev.}\ }\textbf {\bibinfo {volume} {D53}},\
  \bibinfo {pages} {6749} (\bibinfo {year} {1996})},\ \Eprint
  {http://arxiv.org/abs/gr-qc/9511032} {arXiv:gr-qc/9511032 [gr-qc]}
  \BibitemShut {NoStop}%
\bibitem [{\citenamefont {Datta}(2010)}]{datta2010numerical}%
  \BibitemOpen
  \bibfield  {author} {\bibinfo {author} {\bibfnamefont {B.}~\bibnamefont
  {Datta}},\ }\href {http://books.google.co.uk/books?id=-tW8-FUoxWwC} {\emph
  {\bibinfo {title} {Numerical Linear Algebra and Applications}}}\ (\bibinfo
  {publisher} {Society for Industrial and Applied Mathematics},\ \bibinfo
  {year} {2010})\BibitemShut {NoStop}%
\bibitem [{\citenamefont {Berti}\ \emph {et~al.}(2005)\citenamefont {Berti},
  \citenamefont {Buonanno},\ and\ \citenamefont {Will}}]{Berti:2004bd}%
  \BibitemOpen
  \bibfield  {author} {\bibinfo {author} {\bibfnamefont {E.}~\bibnamefont
  {Berti}}, \bibinfo {author} {\bibfnamefont {A.}~\bibnamefont {Buonanno}}, \
  and\ \bibinfo {author} {\bibfnamefont {C.~M.}\ \bibnamefont {Will}},\ }\href
  {\doibase 10.1103/PhysRevD.71.084025} {\bibfield  {journal} {\bibinfo
  {journal} {Phys.Rev.}\ }\textbf {\bibinfo {volume} {D71}},\ \bibinfo {pages}
  {084025} (\bibinfo {year} {2005})},\ \Eprint
  {http://arxiv.org/abs/gr-qc/0411129} {arXiv:gr-qc/0411129 [gr-qc]}
  \BibitemShut {NoStop}%
\bibitem [{\citenamefont {Flanagan}\ and\ \citenamefont
  {Hughes}(1998)}]{Flanagan:1997kp}%
  \BibitemOpen
  \bibfield  {author} {\bibinfo {author} {\bibfnamefont {E.~E.}\ \bibnamefont
  {Flanagan}}\ and\ \bibinfo {author} {\bibfnamefont {S.~A.}\ \bibnamefont
  {Hughes}},\ }\href {\doibase 10.1103/PhysRevD.57.4566} {\bibfield  {journal}
  {\bibinfo  {journal} {Phys.Rev.}\ }\textbf {\bibinfo {volume} {D57}},\
  \bibinfo {pages} {4566} (\bibinfo {year} {1998})},\ \Eprint
  {http://arxiv.org/abs/gr-qc/9710129} {arXiv:gr-qc/9710129 [gr-qc]}
  \BibitemShut {NoStop}%
\bibitem [{\citenamefont {McWilliams}\ \emph {et~al.}(2010)\citenamefont
  {McWilliams}, \citenamefont {Kelly},\ and\ \citenamefont
  {Baker}}]{McWilliams:2010eq}%
  \BibitemOpen
  \bibfield  {author} {\bibinfo {author} {\bibfnamefont {S.~T.}\ \bibnamefont
  {McWilliams}}, \bibinfo {author} {\bibfnamefont {B.~J.}\ \bibnamefont
  {Kelly}}, \ and\ \bibinfo {author} {\bibfnamefont {J.~G.}\ \bibnamefont
  {Baker}},\ }\href {\doibase 10.1103/PhysRevD.82.024014} {\bibfield  {journal}
  {\bibinfo  {journal} {Phys.Rev.}\ }\textbf {\bibinfo {volume} {D82}},\
  \bibinfo {pages} {024014} (\bibinfo {year} {2010})},\ \Eprint
  {http://arxiv.org/abs/1004.0961} {arXiv:1004.0961 [gr-qc]} \BibitemShut
  {NoStop}%
\bibitem [{\citenamefont {Ohme}(2012{\natexlab{b}})}]{Ohme:2011rm}%
  \BibitemOpen
  \bibfield  {author} {\bibinfo {author} {\bibfnamefont {F.}~\bibnamefont
  {Ohme}},\ }\href {\doibase 10.1088/0264-9381/29/12/124002} {\bibfield
  {journal} {\bibinfo  {journal} {Class.Quant.Grav.}\ }\textbf {\bibinfo
  {volume} {29}},\ \bibinfo {pages} {124002} (\bibinfo {year}
  {2012}{\natexlab{b}})},\ \Eprint {http://arxiv.org/abs/1111.3737}
  {arXiv:1111.3737 [gr-qc]} \BibitemShut {NoStop}%
\bibitem [{\citenamefont {Hannam}\ \emph {et~al.}(2010)\citenamefont {Hannam},
  \citenamefont {Husa}, \citenamefont {Ohme}, \citenamefont {M{\"u}ller},\ and\
  \citenamefont {Br{\"u}gmann}}]{Hannam:2010ec}%
  \BibitemOpen
  \bibfield  {author} {\bibinfo {author} {\bibfnamefont {M.}~\bibnamefont
  {Hannam}}, \bibinfo {author} {\bibfnamefont {S.}~\bibnamefont {Husa}},
  \bibinfo {author} {\bibfnamefont {F.}~\bibnamefont {Ohme}}, \bibinfo {author}
  {\bibfnamefont {D.}~\bibnamefont {M{\"u}ller}}, \ and\ \bibinfo {author}
  {\bibfnamefont {B.}~\bibnamefont {Br{\"u}gmann}},\ }\href {\doibase
  10.1103/PhysRevD.82.124008} {\bibfield  {journal} {\bibinfo  {journal}
  {Phys.Rev.}\ }\textbf {\bibinfo {volume} {D82}},\ \bibinfo {pages} {124008}
  (\bibinfo {year} {2010})},\ \Eprint {http://arxiv.org/abs/1007.4789}
  {arXiv:1007.4789 [gr-qc]} \BibitemShut {NoStop}%
\bibitem [{\citenamefont {Schmidt}\ \emph {et~al.}(2011)\citenamefont
  {Schmidt}, \citenamefont {Hannam}, \citenamefont {Husa},\ and\ \citenamefont
  {Ajith}}]{Schmidt:2010it}%
  \BibitemOpen
  \bibfield  {author} {\bibinfo {author} {\bibfnamefont {P.}~\bibnamefont
  {Schmidt}}, \bibinfo {author} {\bibfnamefont {M.}~\bibnamefont {Hannam}},
  \bibinfo {author} {\bibfnamefont {S.}~\bibnamefont {Husa}}, \ and\ \bibinfo
  {author} {\bibfnamefont {P.}~\bibnamefont {Ajith}},\ }\href {\doibase
  10.1103/PhysRevD.84.024046} {\bibfield  {journal} {\bibinfo  {journal}
  {Phys.Rev.}\ }\textbf {\bibinfo {volume} {D84}},\ \bibinfo {pages} {024046}
  (\bibinfo {year} {2011})},\ \Eprint {http://arxiv.org/abs/1012.2879}
  {arXiv:1012.2879 [gr-qc]} \BibitemShut {NoStop}%
\bibitem [{\citenamefont {Boyle}\ \emph {et~al.}(2011)\citenamefont {Boyle},
  \citenamefont {Owen},\ and\ \citenamefont {Pfeiffer}}]{Boyle:2011gg}%
  \BibitemOpen
  \bibfield  {author} {\bibinfo {author} {\bibfnamefont {M.}~\bibnamefont
  {Boyle}}, \bibinfo {author} {\bibfnamefont {R.}~\bibnamefont {Owen}}, \ and\
  \bibinfo {author} {\bibfnamefont {H.~P.}\ \bibnamefont {Pfeiffer}},\ }\href
  {\doibase 10.1103/PhysRevD.84.124011} {\bibfield  {journal} {\bibinfo
  {journal} {Phys.Rev.}\ }\textbf {\bibinfo {volume} {D84}},\ \bibinfo {pages}
  {124011} (\bibinfo {year} {2011})},\ \Eprint {http://arxiv.org/abs/1110.2965}
  {arXiv:1110.2965 [gr-qc]} \BibitemShut {NoStop}%
\bibitem [{\citenamefont {Schmidt}\ \emph {et~al.}(2012)\citenamefont
  {Schmidt}, \citenamefont {Hannam},\ and\ \citenamefont
  {Husa}}]{Schmidt:2012rh}%
  \BibitemOpen
  \bibfield  {author} {\bibinfo {author} {\bibfnamefont {P.}~\bibnamefont
  {Schmidt}}, \bibinfo {author} {\bibfnamefont {M.}~\bibnamefont {Hannam}}, \
  and\ \bibinfo {author} {\bibfnamefont {S.}~\bibnamefont {Husa}},\ }\href
  {\doibase 10.1103/PhysRevD.86.104063} {\bibfield  {journal} {\bibinfo
  {journal} {Phys.Rev.}\ }\textbf {\bibinfo {volume} {D86}},\ \bibinfo {pages}
  {104063} (\bibinfo {year} {2012})},\ \Eprint {http://arxiv.org/abs/1207.3088}
  {arXiv:1207.3088 [gr-qc]} \BibitemShut {NoStop}%
\bibitem [{\citenamefont {Brown}\ \emph
  {et~al.}(2012{\natexlab{b}})\citenamefont {Brown}, \citenamefont {Lundgren},\
  and\ \citenamefont {O'Shaughnessy}}]{Brown:2012gs}%
  \BibitemOpen
  \bibfield  {author} {\bibinfo {author} {\bibfnamefont {D.~A.}\ \bibnamefont
  {Brown}}, \bibinfo {author} {\bibfnamefont {A.}~\bibnamefont {Lundgren}}, \
  and\ \bibinfo {author} {\bibfnamefont {R.}~\bibnamefont {O'Shaughnessy}},\
  }\href {\doibase 10.1103/PhysRevD.86.064020} {\bibfield  {journal} {\bibinfo
  {journal} {Phys.Rev.}\ }\textbf {\bibinfo {volume} {D86}},\ \bibinfo {pages}
  {064020} (\bibinfo {year} {2012}{\natexlab{b}})},\ \Eprint
  {http://arxiv.org/abs/1203.6060} {arXiv:1203.6060 [gr-qc]} \BibitemShut
  {NoStop}%
\bibitem [{\citenamefont {Ajith}\ \emph {et~al.}(2012)\citenamefont {Ajith},
  \citenamefont {Fotopoulos}, \citenamefont {Privitera}, \citenamefont
  {Neunzert},\ and\ \citenamefont {Weinstein}}]{Ajith:2012mn}%
  \BibitemOpen
  \bibfield  {author} {\bibinfo {author} {\bibfnamefont {P.}~\bibnamefont
  {Ajith}}, \bibinfo {author} {\bibfnamefont {N.}~\bibnamefont {Fotopoulos}},
  \bibinfo {author} {\bibfnamefont {S.}~\bibnamefont {Privitera}}, \bibinfo
  {author} {\bibfnamefont {A.}~\bibnamefont {Neunzert}}, \ and\ \bibinfo
  {author} {\bibfnamefont {A.}~\bibnamefont {Weinstein}},\ }\href@noop {} {\
  (\bibinfo {year} {2012})},\ \Eprint {http://arxiv.org/abs/1210.6666}
  {arXiv:1210.6666 [gr-qc]} \BibitemShut {NoStop}%
\bibitem [{\citenamefont {Harry}\ \emph {et~al.}(2013)\citenamefont {Harry},
  \citenamefont {Nitz}, \citenamefont {Brown}, \citenamefont {Lundgren},
  \citenamefont {Ochsner} \emph {et~al.}}]{Harry:2013tca}%
  \BibitemOpen
  \bibfield  {author} {\bibinfo {author} {\bibfnamefont {I.}~\bibnamefont
  {Harry}}, \bibinfo {author} {\bibfnamefont {A.}~\bibnamefont {Nitz}},
  \bibinfo {author} {\bibfnamefont {D.~A.}\ \bibnamefont {Brown}}, \bibinfo
  {author} {\bibfnamefont {A.}~\bibnamefont {Lundgren}}, \bibinfo {author}
  {\bibfnamefont {E.}~\bibnamefont {Ochsner}},  \emph {et~al.},\ }\href@noop {}
  {\  (\bibinfo {year} {2013})},\ \Eprint {http://arxiv.org/abs/1307.3562}
  {arXiv:1307.3562 [gr-qc]} \BibitemShut {NoStop}%
\bibitem [{\citenamefont {P{\"u}rrer}\ \emph {et~al.}(2013)\citenamefont
  {P{\"u}rrer}, \citenamefont {Hannam}, \citenamefont {Ajith},\ and\
  \citenamefont {Husa}}]{Purrer:2013ojf}%
  \BibitemOpen
  \bibfield  {author} {\bibinfo {author} {\bibfnamefont {M.}~\bibnamefont
  {P{\"u}rrer}}, \bibinfo {author} {\bibfnamefont {M.}~\bibnamefont {Hannam}},
  \bibinfo {author} {\bibfnamefont {P.}~\bibnamefont {Ajith}}, \ and\ \bibinfo
  {author} {\bibfnamefont {S.}~\bibnamefont {Husa}},\ }\href@noop {} {\
  (\bibinfo {year} {2013})},\ \Eprint {http://arxiv.org/abs/1306.2320}
  {arXiv:1306.2320 [gr-qc]} \BibitemShut {NoStop}%
\end{thebibliography}%

\end{document}